\documentclass[a4paper,11pt]{article}
\pdfoutput=1 
\usepackage{jheppub} 
\usepackage[T1]{fontenc}
\usepackage[italian,english]{babel}
\usepackage{hyperref}
\hypersetup{
	colorlinks=true,
	linkcolor=[rgb]{0.0,0.42,0.89},
	filecolor=magenta,      
	urlcolor=[rgb]{0.0,0.42,0.89},
	citecolor=[rgb]{0.64,0.0,0.0},
}
\usepackage{ifpdf}
\usepackage{subfigure}
\usepackage{gensymb}
\usepackage{amssymb}
\usepackage{amsfonts}
\usepackage{epsf}
\usepackage{rotating}
\usepackage{graphicx}
\usepackage{amsmath}
\usepackage{fancyhdr}
\usepackage{lineno}
\usepackage{babel, blindtext}
\usepackage{graphics}
\usepackage{color}
\usepackage{physics}
\usepackage{multirow}
\usepackage{pstricks}
\usepackage{xcolor}
\usepackage{multirow}
\usepackage{hhline}
\usepackage{cancel}
\usepackage{framed}
\usepackage{mathtools}
\usepackage{diagbox}
\usepackage{float}
\usepackage[lmargin=1.7in,bmargin=0.2in,footskip=0.5in,total={6.9in,9.5in}]{geometry}

\newcommand{\nn}{\nonumber}
\newcommand{\lsim}{\mathrel{\mathop{\kern 0pt \rlap
  {\raise.2ex\hbox{$<$}}}
  \lower.9ex\hbox{\kern-.190em $\sim$}}}
\newcommand{\gsim}{\mathrel{\mathop{\kern 0pt \rlap
  {\raise.2ex\hbox{$>$}}}
  \lower.9ex\hbox{\kern-.190em $\sim$}}}

\newcommand{\be}{\begin{equation}}
\newcommand{\ee}{\end{equation}}
\newcommand{\bea}{\begin{eqnarray}}
\newcommand{\eea}{\end{eqnarray}}

\newcommand{\bl}{\textcolor{blue}}

\title{\boldmath Exploring CP-violation in $Y=0$ inert triplet with real singlet}

\author[a]{Shilpa Jangid}
\author[a,b,c]{Hiroshi Okada}

\affiliation[a]{
	Asia Pacific Center for Theoretical Physics (APCTP) - Headquarters San 31,
	Hyoja-dong, Nam-gu, Pohang 790-784, Korea}
\affiliation[b]{
	Department of Physics, Pohang University of Science
	and Technology, Pohang 37673, Republic of Korea}
\affiliation[c]{
	Department of Physics, Kyushu University, 744 Motooka, Nishi-ku, Fukuoka, 819-0395, Japan}
\emailAdd{shilpa.jangid@apctp.org, okada.hiroshi@phys.kyushu-u.ac.jp }

\preprint{}

\abstract{In this article, we examine the Standard Model extended with a $Y=0$ Higgs triplet and a real singlet. We consider the Higgs triplet to be odd under the $Z_2$ symmetry, and hence the lightest stable particle from the inert triplet becomes the dark matter candidate, whereas the real singlet is considered to be even under the $Z_2$ symmetry. A dimension-5 effective term is introduced with the help of a real singlet, which breaks the CP symmetry and gives an additional source of CP-violation in the fermion sector. The phase transition proceeds in two-steps, with the symmetry breaking in the singlet direction occurring first and later leading to the usual electroweak symmetry breaking minima, while electroweak baryogenesis is associated with the second step. The parameters chosen for the electroweak phase transition are found to be consistent with the Planck scale stability and the perturbativity using two-loop $\beta$-functions. The DM mass bound for inert triplet, i.e., 1.2 TeV (below which it is under abundance), also comes out to be consistent with the strongly first-order phase transition, which was not possible solely with inert triplet. The upper bound on the triplet mass comes out to be $\leq 3.8$ TeV, which satisfies the strongly first-order phase transition. This particular benchmark point also satisfies the correct baryon asymmetry of the Universe $(6.13 \times 10^{-11})$, and the gravitational wave spectrum also lies within the detectable frequency range of LISA $(6.978 \times 10^{-4} - 1.690 \times 10^{-2} )$ Hz and BBO $(2.80\times 10^{-3}-1.096)$ Hz experiments.}


\begin{document}

\maketitle
\flushbottom

\section{Introduction} 
With the discovery of the Higgs boson, the Standard Model is considered to be the most successful theory so far. This was the last missing piece of the Standard Model (SM) discovered with a mass of 125.5 GeV by the ATLAS \cite{ATLAS:2012yve} and the CMS \cite{CMS:2012qbp} collaborations at the Large Hadron Collider (LHC). Still, there is an obvious shortcoming of the SM in explaining the baryon asymmetry of the Universe. The correct baryon asymmetry, defined as the baryon to entropy ratio $\frac{n_b}{s}\simeq (0.7-0.9)\times 10^{-10}$ \cite{Planck:2013pxb, PhysRevD.86.010001}, cannot be explained in the context of SM. Though the SM accommodate all possible ingredients needed to explain the baryon asymmetry of the Universe \cite{Sakharov:1967dj} i.e., 1) violation of the net baryon number: 2) violation of C-and CP-asymmetry; 3) departure from the thermal equilibrium. The departure from the thermal equilibrium is achieved by a strongly first-order electroweak phase transition, which proceeds via bubble nucleation \cite{PhysRevD.30.2212,Gavela:1994dt,Huet:1994jb}. But, in the case of the SM, the phase transition is not strongly first-order \cite{Aoki:1999fi}, it is indeed a smooth crossover for the Higgs boson mass more than 80 GeV \cite{Kajantie:1996mn,Kajantie:1996qd,Csikor:1998eu}, which is not consistent with the measured Higgs boson mass of 125.5 GeV \cite{Kajantie:1995kf}. The second thing is that the CP violation in the SM provided by the CKM matrix is too small to produce the correct baryon number \cite{Gavela:1993ts,Huet:1994jb,Gavela:1994dt}. Furthermore, the observed dark matter (DM) abundance, i.e., the prediction that the DM constitutes $26 \%$ of our Universe by PLANCK \cite{Planck:2013pxb}and WMAP \cite{Correia:2010jf,WMAP:2012nax}, could not be addressed in the SM, despite the proofs of the existence of the DM from the galaxy rotation curves and from the cosmic microwave background (CMB), etc. These shortcomings indicate the exploration of CP violation beyond the SM \cite{Konstandin:2003dx}.\\

The baryon asymmetry is generated by electroweak baryogenesis (EWBG) during the strongly first-order phase transition  \cite{Trodden:1998ym,Cohen:1993nk,Morrissey:2012db,Baker:2021zsf,Cline:2020jre} satisfying the Sakarov conditions \cite{Sakharov:1967dj}. EWBG and the phase transition are already studied in the context of various SUSY \cite{Carena:1996wj,Delepine:1996vn,Davies:1996qn,Huber:2000mg,Menon:2004wv,Menon:2009mz,Cohen:2012zza,Curtin:2012aa,Carena:2012np,Krizka:2012ah,Delgado:2012eu,Huang:2014ifa,Kozaczuk:2014kva,Baum:2020vfl,Huet:1995sh,Cline:2000nw,Konstandin:2005cd,Riotto:1997vy,Carena:2000id,Chatterjee:2022pxf} and non-SUSY extensions of the SM, i.e., extension with a singlet field \cite{Profumo:2007wc,Espinosa:2011ax,Curtin:2014jma,Jiang:2015cwa,Kurup:2017dzf,Chiang:2017nmu,Kang:2017mkl,Patel:2013zla,Espinosa:1993bs,Espinosa:2011eu}, two-Higgs doublets \cite{Blinov:2015sna,Hambye:2007vf,Ginzburg:2010wa,Jarvinen:2009wr,Gil:2012ya,Chowdhury:2011ga,Borah:2012pu,Cline:2013bln,AbdusSalam:2013eya,Aoki:2021oez,Hammerschmitt:1994fn,Fromme:2006cm}, Higgs triplet \cite{Niemi:2020hto,Patel:2012pi} and many more \cite{Fromme:2006wx,Bodeker:2004ws,Ahriche:2014jna,Bell:2019mbn,Cline:2021dkf}. The electroweak phase transition is also studied with one-step and two-step phase transitions \cite{Patel:2012pi}. Two-step phase transition in the context of singlet has been studied, where the symmetry first breaks in the singlet direction and later, the usual electroweak symmetry breaking is achieved at further lower temperature \cite{Huang:2014ifa,Kozaczuk:2014kva,Chung:2012vg,Ahriche:2007jp,Profumo:2007wc,Curtin:2014jma}. Also, multi-step transition is studied with the exotic doublets \cite{Land:1992sm} in the context of EWBG, where the sphalerons are strongly suppressed in the first transition and this does not work \cite{Hammerschmitt:1994fn}. \\

In this article, we consider the extension of the SM with a real singlet and a $Y=0$ inert triplet with a two-step phase transition. The symmetry first breaks into the singlet direction and then, later, to the usual electroweak broken phase. The singlet field is considered to be even under the $Z_2$ symmetry, and an additional dimensional-5 term involving $\frac{s}{\Lambda}$ is added to generate the baryon number during the course of electroweak phase transition (EWPT) by the additional CP asymmetry. And, the inert triplet, which is odd under the $Z_2$ symmetry provides the much needed DM candidate. The DM candidate from the inert triplet is very heavy in order to satisfy the DM constraints form the relic density, which is not consistent with the strongly first-order phase transition along with the Planck scale perturbativity. There is no possibility for the DM mass from triplet to satisfy the strongly first-order phase transition in case of SM extension solely with the inert triplet. Hence, in this case, it is interesting to see that the possibility of two-step transition using singlet field will help to achieve the correct DM mass consistent with the strongly first-order phase transition and the Planck scale perturbativity and will also provide the correct baryon to entropy ratio.\\

The outline of this work is as follows. The electroweak symmetry breaking (EWSB) in the extension of the SM with the real singlet and the $Y=0$ Higgs triplet along with the possibility of a dimension-5 operator are given in \autoref{model}. The benchmark points that give the measured Higgs boson mass are tested with the bounds from the Planck scale vacuum stability and the perturbativity using two-loop $\beta-$functions in \autoref{stability}. In \autoref{potential}, the effective potential at finite temperature is discussed along with the thermal masses. The two-loop contributions to the thermal masses using the dimensional reduction method are also discussed. The two-step EWPT is discussed in \autoref{ewpt}. The phase transition proceeds via developing a vacuum expectation value (vev) along the singlet direction first and then to the usual electroweak mimina, and the EWBG is to be achieved during the second step, which is discussed in \autoref{EWBG}. The gravitational wave signatures (GW) arising from the strongly first-order phase transition are explored in \autoref{gw}. Eventually, the conclusions are given in \autoref{conc}. The expressions for the two-loop $\beta-$functions, the dimensional reduction calculation, and the transport equations for the baryogenesis are given in \autoref{betaf1}, \autoref{3dexp} and \autoref{trans}, respectively.

\section{ITM plus real singlet}\label{model}
The minimal SM is extended with a $Y=0$ Higgs triplet and a real singlet. The Higgs triplet is considered odd under the discrete $Z_2$ symmetry and does not take part in the EWSB, termed the "inert triplet model (ITM)". The real singlet, being even under the $Z_2$ symmetry, acquires vev and provides an additional source of CP-violation through higher dimension operator. The $Z_2$ symmetry assignment for all the fields is as follows:
\bea
Z_2:  \Phi \rightarrow \Phi, \Delta \rightarrow -\Delta, S \rightarrow S,
\eea
where,
\begin{center}
	$	\Phi
	= \left(\begin{array}{c}
		G^+   \\
		\frac{1}{\sqrt{2}}(v+\rho_1+i G^0)  \end{array}\right) $, \qquad \qquad
	$\Delta =\frac{1}{2} \left(
	\begin{array}{cc}
		\Delta^0 & \sqrt{2} \Delta^+ \\
		\sqrt{2} \Delta^- & -\Delta^0 \\
	\end{array}
	\right)$, \qquad \qquad $S=\frac{1}{\sqrt{2}}(x  + \rho_2 ).$
\end{center}
Since $S$ is even under the $Z_2$ symmetry, mixing is allowed only with the neutral component of $\Phi$ and the singlet vev $x$ is related to the SM vev as $\tan\beta = \frac{x}{v}$.
The full scalar sector of this model is described below:
\bea
V_0= V_{ITM}+V_S + V_{HTS},\label{Eq:2.2} 
\eea
where, the scalar potential for the inert triplet model is given as:
\be
 V_{ITM}=m_h^2  \Phi^\dagger \Phi+m_\Delta^2  Tr(\Delta^\dagger \Delta)+\lambda_h|\Phi^\dagger \Phi|^2+\lambda_\Delta(Tr|\Delta^\dagger \Delta|)^2+\lambda_{h\Delta}\Phi^\dagger \Phi Tr(\Delta^\dagger \Delta)
 +\lambda'_{h\Delta}\sum_{i=1}^3\Phi^\dagger \sigma_i\Phi Tr(\Delta^\dagger\sigma_i \Delta),\\ \label{Eq:2.3} 
\ee
where $\sigma_i(i=1-3)$ are Pauli matrices and $\lambda'_{h\Delta}$ does not affect the RGE thus we neglect this term hereafter.
The scalar potential for the singlet is written as:
\bea
 V_S & = & m_{S}^2  S^2 + \lambda_{S}S^4 
  +\alpha_1  S+\kappa S^3, \label{Eq:2.4}
\eea
where the linear term can be eliminated by a translation of $S$, i.e., $\alpha_1$.
And the corresponding interaction terms between the doublet-singlet and the triplet-singlet are as follows:
\bea
V_{HTS}& = & \lambda_{hs}(\Phi^\dagger\Phi)S^2 +   \lambda_{\Delta s} Tr(\Delta^\dagger \Delta) S^2 
 + \alpha_2 (\Phi^\dagger \Phi)S + \alpha_3  Tr(\Delta^\dagger \Delta)S.
\eea
 If we choose singlet field to be also odd under the $Z_2$ symmetry, all the odd terms in field $S$ will be eliminated. But the tree-level barrier generated with the singlet cubic term is crucial for first-order phase transition and hence, we are not forcing the singlet to be the odd under the $Z_2$ symmetry. Hence, we consider only the terms that are cubic in the singlet field $\kappa$, neglecting the remaining ones $\alpha_{2,3}$. The lightest stable particle from the inert triplet becomes the DM candidate, and $\Phi$ is the portal for the DM interactions with the visible sector. Therefore, we assume that there is no direct coupling of $S$ to the triplet field and only possibility of $S$ to interact with the DM particles is via the mixing with the neutral component of $\Phi$ \cite{Bonilla:2014xba}. As a result, the direct coupling terms between the singlet and the triplet, which are $Z_2$ symmetric $(\lambda_{\Delta s})$ are still assumed to be zero. 
 After all these assumptions, the full scalar potential in \autoref{Eq:2.2} is rewritten as:
 \bea
 V_0 = m_h^2  \Phi^\dagger \Phi+m_\Delta^2  Tr(\Delta^\dagger \Delta)+\lambda_h|\Phi^\dagger \Phi|^2+\lambda_\Delta(Tr|\Delta^\dagger \Delta|)^2+\lambda_{h\Delta}\Phi^\dagger \Phi Tr(\Delta^\dagger \Delta) +m_{S}^2  S^2 + \lambda_{S}S^4 
 + \kappa S^3+ \lambda_{hs}(\Phi^\dagger\Phi)S^2. \nn \\ 
 \label{eq:2.6}
 \eea
 Since singlet has no direct coupling to the gauge bosons and fermions in the SM, the $G^+$ and $G^0$ in the Higgs doublet ($\Phi$) are the massless Goldstone bosons, which provide mass to the $W^{\pm}$ and $Z$ bosons. The minimization conditions after the EWSB are computed as follows:
 \bea
 m_h^2 & = & -\lambda_h v^2 - \frac{1}{2}\lambda_{hs}x^2, \nn \\
 m_S^2 & = & -\frac{3}{2\sqrt{2}} \kappa x - \frac{1}{2}\lambda_{hs} v^2 - \lambda_s x^2.
 \label{eq:2.7}
 \eea
 The neutral component of singlet mixes with the SM doublet, and the gauge eigenstates $(\rho_1,\rho_2)$ are rotated to the mass eigenstates with the rotation matrix as given below:
 \begin{center}
 	$	 \left(\begin{array}{c}
 		h   \\
 		H
 	 \end{array}\right) =
  \left(
 	\begin{array}{cc}
 		\cos\theta & -\sin\theta \\
 		\sin\theta & \cos\theta \\
 	\end{array}
 	\right) \left(\begin{array}{c}
 	\rho_1   \\
 	\rho_2\end{array}\right),$
 \end{center}
and the squared mass mixing matrix for neutral Higgs is given as:
 \begin{center}
	$ \mathcal{M}^2 =	 
	\left(
	\begin{array}{cc}
		2  \lambda_h v^2  & \lambda_{hs}v x \\
		\lambda_{hs}v x & 2(\frac{3 \kappa x}{4 \sqrt{2}}+\lambda_s x^2)\\
	\end{array}
	\right)= \left(
	\begin{array}{cc}
		A  & B \\
		B & C\\
	\end{array}
	\right).$
\end{center}
After diagonalization, we obtain the corresponding mass eigenstates for neutral scalars as follows:
\bea
M^2_h =  (A+C) - \sqrt{(A-C)^2+B^2}, \nn \\
M^2_H = (A+C) + \sqrt{(A-C)^2+B^2}.  
\eea
where, $h$ is identified as the SM Higgs boson with mass of 125.5 GeV. The constraint on the mixing angle $-\frac{\pi}{2}\leq \theta \leq \frac{\pi}{2}$ is defined as;
\bea
\sin 2 \theta = \frac{B}{\sqrt{(A-C)^2+B^2}}, \nn \\
\cos 2 \theta = \frac{C-A}{\sqrt{(A-C)^2+B^2}}. 
\eea
 The value of the mixing angle $\sin\theta$, is bounded by both theoretical and experimental constraints as $|\sin\theta | < 0.33$ \cite{Profumo:2007wc}. The mass eigenstates for triplet will be the same as the gauge eigenstates as it does not take part in the EWSB, and the mass expressions are as follows:
\bea
\label{eq:2.3}
	M_{\Delta^0}^2  & = & m_\Delta^2 + \frac{\lambda_{h\Delta}}{2}v^2,  \\
	M_{\Delta^{\pm}}^2  & = & m_\Delta^2 + \frac{\lambda_{h\Delta}}{2}v^2.
\eea
 Note that both neutral mass eigenstates $(h,H)$ are CP-even, and there is no possibility of introducing the CP-violation, neither explicitly by introducing complex parameters into the potential nor spontaneous CP-violation by the complex vev of the singlet. The relevant Yukawa interactions for the leptons, the up-type quarks, and the down-type quarks are given as follows:
 \begin{eqnarray}
 	\mathcal{L}_{Yukawa} = - \overline{Q_{L}^{}}\Phi Y_{d} d_{R}^{} - \overline{Q_{L}^{}}\widetilde{\Phi_{}}Y_{u}^{}  u_{R}^{} - \overline{L_{L}^{}}\Phi_{}Y_{l} l_{R}^{} + H.c.,
 \end{eqnarray}
 where $\widetilde{\Phi}= i\sigma_2\Phi^*$, $Q_L$ and $L_L$ are the left-handed quark and lepton doublets under $SU(2)$ and $d_R, u_R$, and $e_R$ are the right-handed singlets under $SU(2)$, respectively. The only source of CP-violation is in the fermionic sector, which is introduced by introducing a higher dimension operator. The corresponding dimension-5 effective operator in the absence of $Z_2$ symmetry is written as:
\bea
\mathcal{O}_5= \frac{\alpha }{\Lambda}S\overline{Q}_{3L}\widetilde{\Phi}t_R + H.c.,\label{eq:2.13}
\eea
where, $\alpha$ is the complex CP-violating parameter.
Here, $\overline{Q}_{3L}$ and $t_R $ denote only the third-generation for left-handed quark doublet and right-handed quark singlet fields, and $\Lambda$ is the cut-off scale parameterizing the amplitude of the effective operator. 

In the next section, we discuss the constraints imposed by the vacuum stability and the Planck scale perturbativity on the parameter space of the model.

\section{Vacuum stability and perturbativity constraints}\label{stability}
In this section, we discuss the stability of the vacuum and the perturbative unitarity constraints up to Planck scale. In order to achieve the stability of the electroweak vacuum, the scalar potential must be bounded from below. The tree-level conditions ensuring the scalar potential to be stable are as follows \cite{Araki:2010nak}:
\bea
\lambda_{h}\geq 0, \quad \lambda_{\Delta} \geq 0, \quad \lambda_s \geq 0, \quad \lambda_{h\Delta}\geq -\sqrt{\lambda_h \lambda_{\Delta}}, \quad \lambda_{hs} \geq -\sqrt{\lambda_h \lambda_s},
\quad  0(=\lambda_{\Delta s}) \geq -\sqrt{\lambda_\Delta \lambda_s}.
\eea
The dimensionless couplings are constrained to certain values to ensure the perturbative bounds reach a particular energy scale $\mu$. The constraints on the dimensionless couplings from perturbative unitarity are as follows; \cite{Jangid:2020dqh,Bandyopadhyay:2020djh}:
\begin{align}
	\left|\lambda_i\right|  \ \leq \ 4 \pi, \qquad
	\left|g_j\right| \ \leq \ 4 \pi, \qquad \left|Y_k\right|  \ \leq \ \sqrt{4\pi} \, .\label{3.20}
\end{align}
Here, $g_j$ with $j=1,2,3$ and $Y_k$ with $k=u,d,\ell$ are the EW gauge couplings and the Yukawa couplings for up- type quarks, down type quarks, and leptons, respectively. The quartic couplings $\lambda_i$ correspond to $ \lambda_h, \lambda_s, \lambda_{hs}, \lambda_{\Delta}, \lambda_{h\Delta}$. The allowed benchmark points for EWSB, vacuum stability, and Planck scale perturbativity are given in \autoref{tab:table1}.

\begin{table}[h]
	\begin{center}
		\begin{tabular}{|c|c|c|c|c|c|c|}\hline
			& 	$M_H $ (GeV)  & $\tan\beta$ & $\lambda_s$ & $\lambda_{hs}$  & 	$\kappa$[GeV]  &  $\lambda_{h}$ \\ \hline
				BP1	 & 177.26 & 1.62  & 0.098 & 0.01 & -0.069 & 0.13 \\  \hline
		BP2	& 177.26  & 1.62  & 0.098 & 0.01 & -0.019 & 0.13\\  \hline	
		\end{tabular}
	\end{center}
	\caption{The chosen benchmark points at the electroweak (EW) scale allowed from symmetry breaking, vacuum stability, and Planck scale perturbativity. The bounds on the quartic couplings from Planck scale perturbativity for the triplet, i.e., $\lambda_{h\Delta}$ and $\lambda_{\Delta}$, are fixed to 1.25 and 0.04, respectively. }	\label{tab:table1}
\end{table}

\begin{figure}[h!]
		\hspace{-0.2cm}
		\mbox{
			\subfigure[Stablity]{\includegraphics[width=0.5\linewidth,angle=-0]{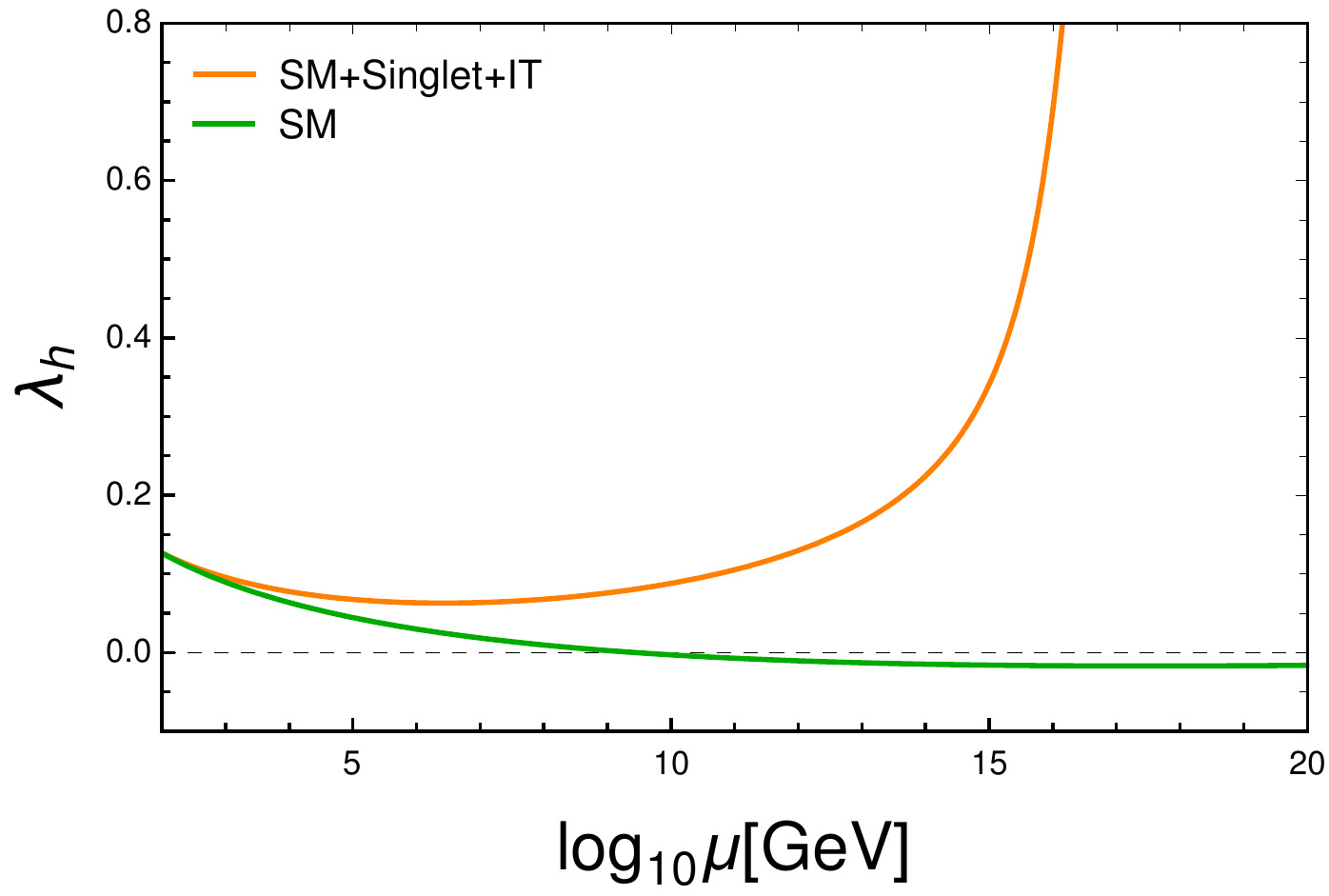}}
			\subfigure[Perturbativity]{\includegraphics[width=0.5\linewidth,angle=-0]{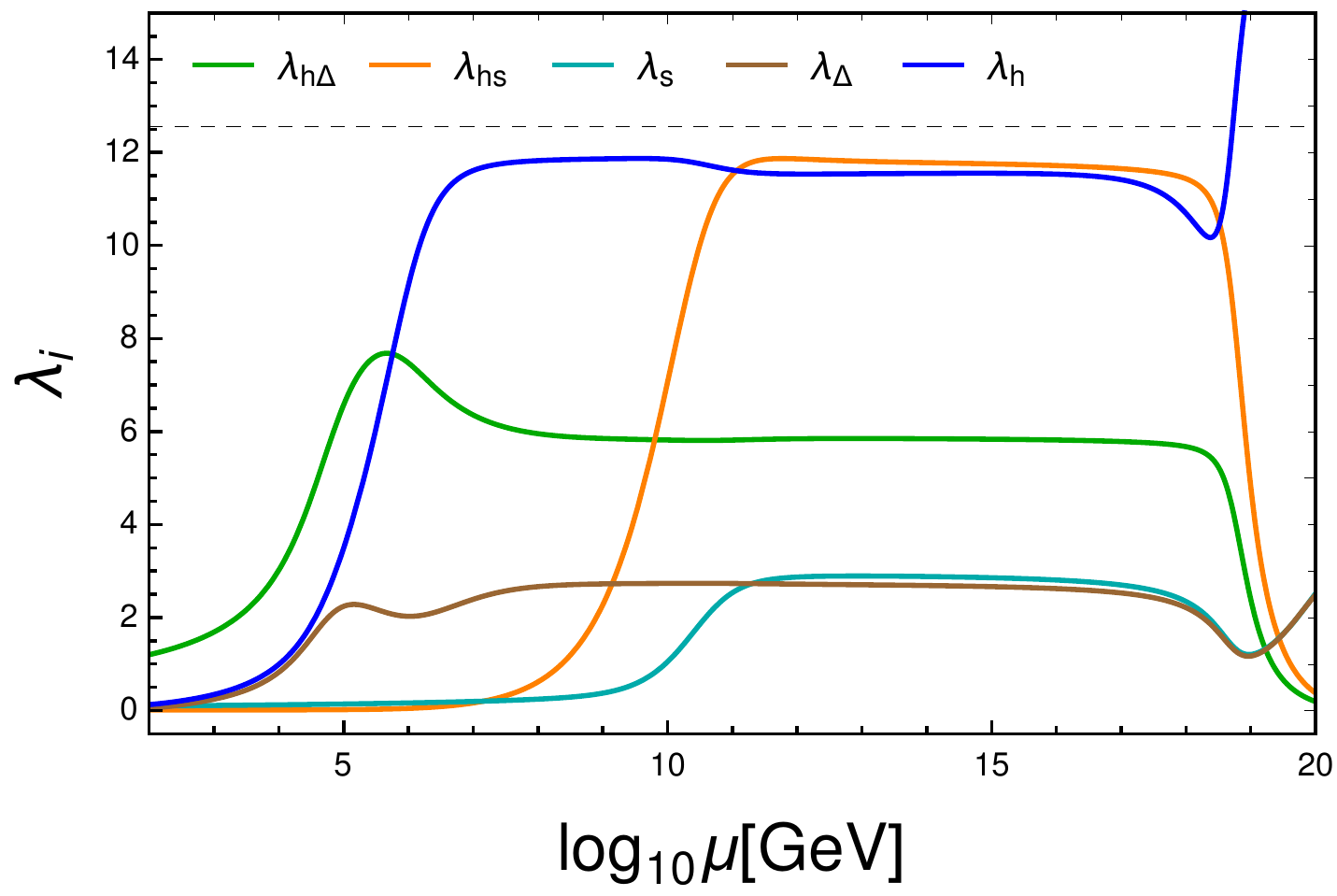}}}
		\caption{(a) Running of Higgs quartic coupling with the energy scale in GeV for stability using two-loop $\beta-$functions with green and orangle curve for SM and SM+Singlet+IT scenario, respectively for $\lambda_{h\Delta}=0.30$ and $\lambda_\Delta=0.04$ at the EW scale; (b) the variation of other dimensionless quartic couplings with the energy scale in GeV for perturbative unitarity till Planck scale for BP2 and the maximum values allowed for $\lambda_\Delta$ and $\lambda_{h\Delta}$ at the EW scale are 0.04 and 1.25, respectively.
		}\label{fig1l}
\end{figure}

The running of the dimensionless couplings is computed using {\tt SARAH} \cite{Staub:2013tta,Staub:2015kfa}, and the full two-loop $\beta-$ functions for the scalar quartic couplings and the gauge couplings are given in \autoref{betaf1}.
The variation of the Higgs quartic coupling $\lambda_h$ and other quartic couplings $\lambda_i \in \lambda_{h\Delta},\lambda_{hs}, \lambda_s, \lambda_\Delta$ with the energy scale $\mu$ are given in~\autoref{fig1l} using BP2. The Higgs quartic coupling becomes negative around $10^9$ GeV in case of the SM \cite{Degrassi:2012ry,Buttazzo:2013uya,Elias-Miro:2011sqh}, as manifested by the green curve in \autoref{fig1l}(a), and this stability scale increases for SM+Singlet+IT till Planck scale with the addition of extra scalar degrees of freedom as delineated by the orange curve in \autoref{fig1l}(a) for $\lambda_{h\Delta}=0.30$ and $\lambda_{\Delta}=0.04$ at the EW scale. Additionally, for the chosen benchmark points, the dimensionless couplings for the theory also satisfy perturbative unitarity up to Planck scale. The $\lambda_{h\Delta}$ coupling is chosen to be the maximum allowed value at the electroweak scale for which Planck scale perturbativity can be achieved, i.e., 1.25 and $\lambda_{\Delta}=0.04$ in \autoref{fig1l}(b).

  After discussing the stability of the vacuum and the perturbative unitarity, we are going to discuss the EWPT from the symmetric phase at high temperature to the broken phase. The expression for the finite temperature effective potential at one-loop and the corresponding thermal corrections to the zero-temperature masses are also discussed in detail in the next section.  
 
\section{Effective potential at finite temperature}\label{potential}
In the case of the SM and its extensions, a cubic term is generated in the Higgs scalar potential by the thermal effects of the bosons coupled to the Higgs, and this cubic term actually triggers the first-order EWPT. This can be achieved in several ways: sizable couplings of these bosons to the Higgs are needed and the effect can then be screened by thermal masses, considering Daisy resummation into account. Here, we consider the EWPT, where the barrier between the symmetric and the broken phase is enhanced by the tree-level effects along with the thermal cubic correction. For tree-level effects, the Higgs vev at critical temperature $v_{c}$ is independent of the temperature and is proportional to some dimensionful parameters in the potential, which leads to strong EWPT, i.e., $\frac{v_{c}}{T_c}$  becomes potentially very large even for lower $T_C$. Therefore, when there exists a large barrier between the electroweak breaking vacuum  $\expval{\Phi}=\frac{v}{\sqrt{2}}$ and $\expval{S}=\frac{x}{\sqrt{2}} $ and a nearly degenerate symmetric one with $\expval{\Phi}=0$, $\expval{S}=\frac{x0}{\sqrt{2}}$, the phase transition can be achieved easily by weak thermal corrections, i.e., at a significantly lower temperature compared to the Higgs vev at the critical temperature $v_{c}$, which can be very close to the vev at zero temperature $v$. The tree-level potential in \autoref{eq:2.6} can be expressed in terms of the background fields $h_1$, $h_2$ and $h_3$ as follows:
\bea
V_0^{\rm eff}= \frac{\kappa h_3^2}{2 \sqrt{2}} + \frac{1}{2}m_h^2 h_1^2 +\frac{1}{2}m_S^2 h_3^2 +\frac{1}{2}m_{\Delta}^2h_2^2 +\frac{1}{4} \lambda_h h_1^4 +\frac{1}{4} \lambda_{hs}h_1^2 h_3^2  +\frac{1}{4}\lambda_{h\Delta}h_1^2 h_2^2 + \frac{1}{4}\lambda_s h_3^4 + \frac{1}{4}\lambda_\Delta h_2^4,
\label{eq:4.1}
\eea
with the following convention;
\begin{center}
	$	\Phi
	= \left(\begin{array}{c}
		0   \\
		\frac{h_1}{\sqrt{2}}  \end{array}\right) $, \qquad \qquad
	$\Delta =\frac{1}{2} \left(
	\begin{array}{cc}
		h_2 & 0 \\
	0 & -h_2 \\
	\end{array}
	\right)$, \qquad \qquad $S=\frac{h_3}{\sqrt{2}}.$
\end{center}

In order to make sure that the tree-level scalar potential has the stationary point at the physical minimum, $m_h^2$ and $m_S^2$ are replaced by equations given in \autoref{eq:2.7}. The one-loop effective potential contribution at zero temperature is given by the Coleman-Weinberg potential and is given as \cite{Coleman}:
\bea
V_{1-loop}^{\rm CW}=\frac{1}{(64\pi)^2}\sum_{i=B,F}(-1)^{F_i} n_i \hat{m_i}^4\Big[\log\Big(\frac{\hat{m}_i^2}{\mu^2}\Big)-k_i\Big],
\eea
where, $F_i =0$ and $1$ for the boson and the fermions, respectively. The constant $k_i$ comes out to be $3/2$ for scalars, longitudinally polarized vector bosons, and the fermions, while $k_i=1/2$ for transverse vector bosons
 and $\hat{m}_i^2$ is the field dependent mass, i.e., $\hat{m}_i^2(h_1,h_2,h_3)$ computed from $V_0^{\rm eff}$ in \autoref{eq:4.1}. Including the one-loop Coleman-Weinberg contributions, the effective scalar potential at zero temperature now becomes;
 \bea
 V_1(T=0) = V_0^{\rm eff} + V_{\rm 1-loop}^{\rm CW}.
 \eea
 After discussing the effective scalar potential at zero temperature, the finite temperature corrections to this potential have to be taken into account. The one-loop potential at a finite temperature is computed as follows;
\bea\label{eq:4.4}
V_{1-loop}^{T \neq 0} =\frac{T^4}{(2\pi)^2}\sum_{i=B,F} (-1)^{F_i} n_i J_{B/F}\Big(\frac{\widetilde{m}_i^2}{T^2}\Big),
\eea
where, $n_i$ accounts for the degrees of freedom of species $i$, and $\widetilde{m}_i^2$ is the thermally corrected mass, which includes contributions from the Daisy corrections resuming hard thermal loops;
\bea
\widetilde{m}_i^2 = \widetilde{m}_i^2(h_1, h_2, h_3; T) = \hat{m}_i^2(h_1, h_2, h_3) + \Pi_i T^2,
\eea
where, $\Pi_i's$ are the Daisy coefficients, which are non-zero only for the bosonic fields. Furthermore, out of vector bosons, only longitudinal polarization states acquire non-zero Daisy corrections, while the gauge symmetry protects the transverse states from any corrections. In general, the expressions for the spline functions $J_{B,F}$ are defined as;
\bea
J_{B,F}(x^2) = \int_{0}^{\infty} dy y^2 \log(1 \mp e^{-\sqrt{y^2+x^2}}).
\eea
The full one-loop finite temperature effective potential can now be given as;
\bea \label{eq:4.7}
V_1(T)= V_0^{\rm eff} + V_{\rm 1-loop}^{\rm CW}(\widetilde{m}_i^2) + V_{\rm 1-loop}^{T \neq 0} (\widetilde{m}_i^2).
\eea
In pursuance of studying the vacuum structure at very high temperatures, i.e., $T^2  >> \hat{m}_i^2$, the finite temperature potential after neglecting the Daisy coefficients can be rewritten as follows;
\bea
V_{\rm 1-loop}^{T\neq 0} \rightarrow T^4[...] + \frac{T^2}{48}\Big(2 \sum_{i=B} n_i \hat{m}_i^2 + \sum_{i=F} n_i \hat{m}_i^2 \Big) + T^4 \times \mathcal{O}\Big(\Big|\frac{\hat{m}_i^2}{T^2}\Big|^{3/2}\Big).
\eea

The ellipsis [....] in the above equation indicates those terms that are independent of field values. The Daisy coefficients $\Pi_i$ can be computed from the one-loop thermal potential in the high temperature limit as given below;
\bea
\Pi_{ij}=\frac{1}{T^2} \frac{\partial^2 V_{\rm 1-loop}^{T \neq 0}(\hat{m}^2)}{\partial \phi_i \partial \phi_j} \Big|_{T>> \hat{m}^2}.
\eea
It is important to note that the Daisy coefficients are computed using field dependent masses, i.e., $\hat{m}_i^2$, which are temperature independent, and later, Daisy-resummed thermal masses $\widetilde{m}_i^2$ are inserted back in $V_{\rm 1-loop}^{T \neq 0}$ as well as in the zero temperature Coleman-Weinberg potential while computing the full temperature-dependent effective potential. The field dependent masses which contribute to the zero temperature effective potential are given as; 
\bea
\hat{m}_{h^2} = \left(
\begin{array}{cc}
	3\lambda_h h_1^2 + m_h^2 + \frac{1}{2}\lambda_{hs}h_3^2 & \lambda_{hs}h_1 h_3 \\
	 \lambda_{hs}h_1 h_3 &  \frac{3}{\sqrt{2}}\kappa h_3 + \frac{1}{2}\lambda_{hs}h_1^2+ m_S^2 + 3 \lambda_s h_3^2\\
\end{array}
\right), 
\eea
\bea
\hat{m}_{G^0}^2 = \lambda_h h_1^2 - m_h^2, \qquad \hat{m}_{W}^2 = \frac{g_2^2}{4} h_1^2, \qquad \hat{m}_{Z}^2 = \frac{g_2^2 + g_1^2}{4}h_1^2, \qquad \hat{m}_t^2 =\frac{y_t^2}{2}h_1^2, \nn
\eea
where, $\hat{m}_{G_0}, \hat{m}_{W}^2, \hat{m}_Z^2, \hat{m}_t^2$ are the masses for the Goldstone bosons, the gauge bosons, and the top quark, respectively. Since the triplet field does not acquire any vev, the corresponding field dependent masses for the triplet will be in terms of the background field of the SM doublet only, i.e., $h_1$
and the corresponding mass expressions for the neutral and the charged component of the triplet field are given as;
\bea
\hat{m}_{\Delta^0}^2 = m_{\Delta}^2 + \frac{\lambda_{h\Delta}}{2}h_1^2, \nn  \\
\hat{m}_{\Delta^{\pm}}^2 =  m_{\Delta}^2 + \frac{\lambda_{h\Delta}}{2}h_1^2.
\eea
The degrees of freedom $n_i$ used in \autoref{eq:4.4} for the SM fields, singlet fields and the triplet fields are given as;
\bea
n_h = 1, n_H=1, n_{G}=3, n_\Delta =3, n_t =12, \nn \\
n_{W_L}=n_{Z_L}=n_{\gamma_L}=1, n_{W_T}=n_{Z_T}=n_{\gamma_T}=2,
\eea
 and the corresponding Daisy coeffiecients for the bosonic degrees of freedom are computed as;
 \bea
\Pi_h & = & \Big(\frac{g_1^2+3g_2^2}{16} + \frac{\lambda_h}{2}+ \frac{y_t^2}{4}+ \frac{\lambda_{h\Delta}}{8}+\frac{\lambda_{hs}}{24}\Big)T^2, \nn \\
 \Pi_G & = & \Big(\frac{g_1^2+3g_2^2}{16} + \frac{\lambda_h}{2}+ \frac{y_t^2}{4}+ \frac{\lambda_{h\Delta}}{8}+\frac{\lambda_{hs}}{24}\Big)T^2, \nn \\
 \Pi_H & = & \Big(  \frac{\lambda_{hs}}{24} + \frac{\lambda_s}{4}\Big)T^2, \nn \\
 \Pi_\Delta & = & \Big(\frac{\lambda_{h\Delta}}{24}+\frac{3\lambda_\Delta}{4}\Big) T^2, \nn \\
 \Pi_{W_L} & = & \frac{11}{6}g_2^2 T^2, \nn \\
 \Pi_{W_T} & = & \Pi_{Z_T}=\Pi_{\gamma_T} =0.
 \eea 
As mentioned previously, only longitudinal components of the gauge bosons, i.e., $W_L, Z_L$, and $\gamma_L$ receive self energy contributions, while the Daisy corrections are zero for the transverse components of the gauge bosons. The thermally corrected mass expressions for the longitudinal components of the $Z_L$ boson and the photon, $\gamma_L$, are given as \cite{Comelli:1996vm};
\bea
\widetilde{m}_{Z_L}^2 & = & \frac{1}{2}\Big[\hat{m}_Z^2 +\frac{11}{16}\frac{g_2^2}{\cos^2\theta_W}T^2 + \delta\Big], \nn \\
\widetilde{m}_{\gamma_L}^2 & = & \frac{1}{2}\Big[\hat{m}_Z^2 +\frac{11}{16}\frac{g_2^2}{\cos^2\theta_W}T^2 - \delta\Big], \nn \\
\eea

where, $\delta$ is given as;
\bea
\delta^2 = \hat{m}_Z^4 + \frac{11}{3} \frac{g_2^2 \cos^2\theta_W}{\cos^2\theta_W}\Big[\hat{m}_Z^2 + \frac{11}{12}\frac{g_2^2}{\cos^2\theta_W}T^2\Big]T^2.
\eea

This section completes the derivation of one-loop effective potential at finite temperature. But, as we are using two-loop $\beta$-functions for the running of the couplings with the energy scale, the inclusion of two-loop corrections specifically at finite-temperature is very important. And the details for the two-loop corrections are given in the next section.

\subsection{Dimensional reduction}
The finite temperature effective potential has residual scale dependence at $\mathcal{O}(g^4)$. In order to cancel this renormalization scale dependence, the explicit logarithms of the renormalization scale are needed, which are achieved at the two-loop level. The two-loop corrections to the thermal masses actually depend on the explicit logarithms of the renormalization scale and cancel out this scale dependence at $\mathcal{O}(g^4)$. These two-loop corrections are computed using the high-temperature dimensional reduction from 4d to a three-dimensional effective theory (3d EFT), and in this reduced 3dEFT, all the parameters in the theory becomes renormalization scale dependent \cite{Gould:2021oba,Farakos:1994kx,PhysRevD.23.2916,Croon:2020cgk}. The expressions for the parameters of the theory are given in detail in \autoref{3dexp}. 

\section{Electroweak phase transition (EWPT)}\label{ewpt}
\begin{figure}[H]
	\begin{center}
		\includegraphics[width=0.6\linewidth,angle=-0]{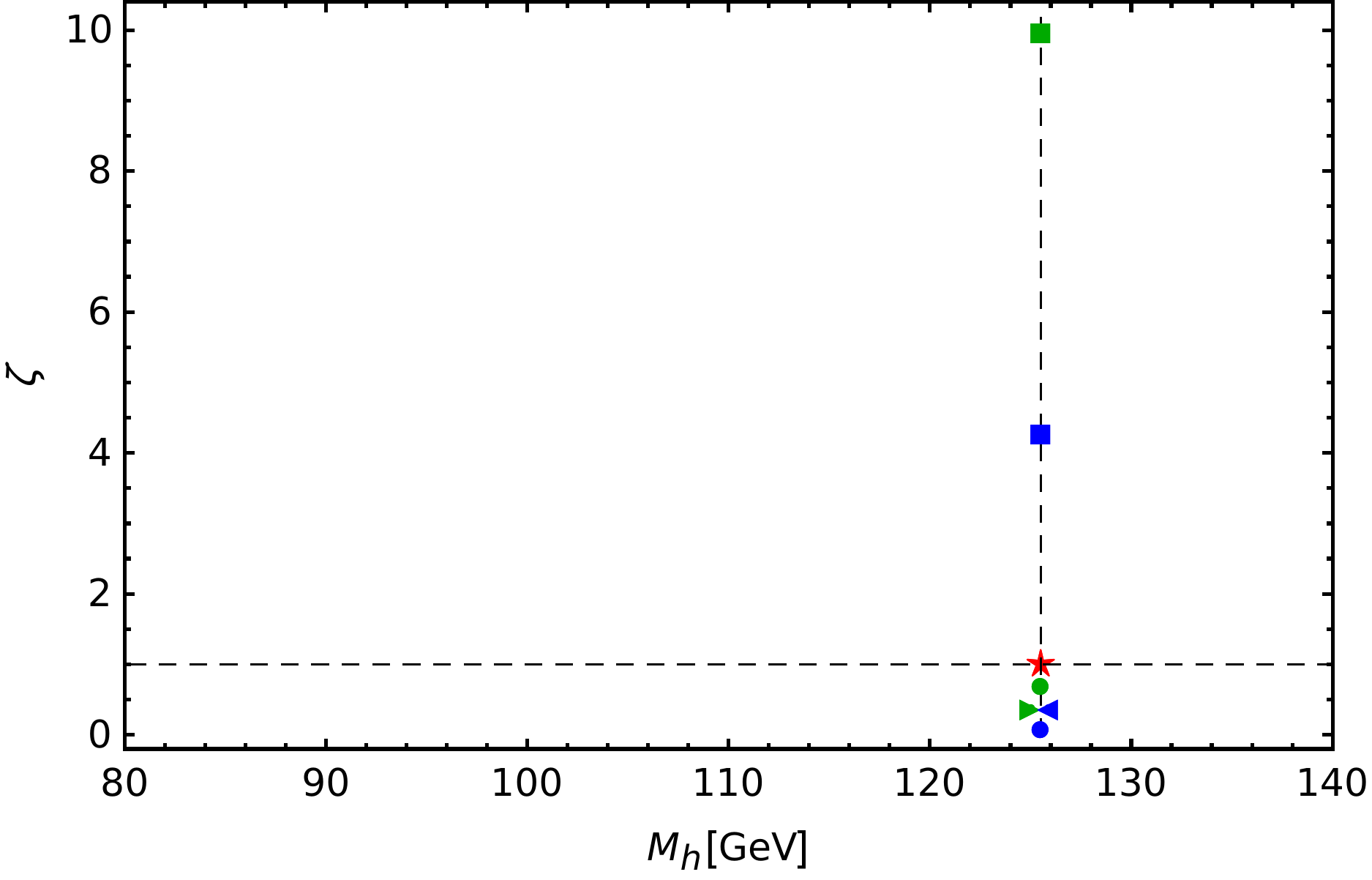}
		\caption{Variation of the order of phase transition $(\zeta=\frac{h_c}{T_c})$ with the Higgs boson mass in GeV. The red star corresponds to the measured Higgs boson mass of 125.5 GeV. The color coding from green to blue corresponds to two different values of the bare mass parameter for the triplet field 700 GeV and 1200 GeV. The different transitions, i.e., Case 1: $V_1^{\rm symm}(0,0) = V_1(\sqrt{-m_h^2/\lambda_h},0)$, Case 2: $V_1^{\rm symm}(0,0) = V_1(0, \sqrt{-m_S^2/\lambda_S})$, and  Case 3: $V_1(\sqrt{-m_h^2/\lambda_h},0)=V_1(0, \sqrt{-m_S^2/\lambda_S})$ for computing the order of phase transition are depicted by different plotmarkers, i.e., circle, triangle and square, respectively.
		}\label{fig2l}
	\end{center}
\end{figure}

After computing the two-loop thermal corrections to the thermal masses and including the bounds from Planck scale perturbative unitarity using two-loop $\beta$-functions, we can proceed with the EWPT from the symmetric phase to the broken phase. Since the triplet field is odd under the $Z_2$ symmetry, there is no symmetry breaking along this direction. Hence, the triplet masses are in terms of the Higgs field itself. In order to study the phase structure at different temperatures, we need to compute the one-loop effective potential in \autoref{eq:4.7} in different phases as given below \cite{Niemi:2020hto};
\bea
V_1^{\rm symm}(T)=V_1(0,0), \qquad \qquad V_1^{\Phi}(T)=V_1(\sqrt{-m_h^2/\lambda_h},0), \qquad \qquad V_1^{S}(T)=V_1(0, \sqrt{-m_S^2/\lambda_S}),
\eea
and then varying the temperature, we need to check for the phases which exist simultaneously. The condition for the critical temperature $T_c$ in this case will be the temperature where the value of one-loop effective potential in any two phases or minima is degenerate. For example: $V_1^{\Phi}(T_c)=V_1^{S}(T_c)$ for $\Phi \rightarrow S$ transitions.
\autoref{fig2l} describes the variation of the order of phase transition $\zeta=\frac{h_c}{T_c}=\frac{\sqrt{h_1(T_c)+h_2^2(T_c)+(h_{3}(T_c)-h_{3}^h(T_c))^2}}{T_c}$ (in case of multiplets \cite{Ahriche:2014jna,Ahriche:2007jp}, when the false vacuum is $(0.0,0.0,h_{3}^h)$ instead of $(0.0,0.0,0.0)$ in the singlet direction) with the Higgs boson mass in GeV. The vev for the triplet field is considered as zero at all possible temperatures, hence, we are left with the minima along the Higgs direction $(h_1)$ and the singlet field direction $(h_3)$. The occurance of a treel-level saddle point between the $\Phi$ and $S$ minima provides a first-order phase transition for $(S \rightarrow \Phi)$ transition.  The red star corresponds to the point that satisfies the measured Higgs boson mass and the criteria for a strongly first-order phase transition. The blue and the green colors correspond to the bare mass parameter for the Higgs triplet, i.e., $m_\Delta$= 700 GeV and 1200 GeV, respectively, keeping the other parameters fixed from BP2 in \autoref{tab:table1}. The different transitions are as follows: Case 1: $V_1^{\rm symm}(0,0) = V_1(\sqrt{-m_h^2/\lambda_h},0)$, Case 2: $V_1^{\rm symm}(0,0) = V_1(0, \sqrt{-m_S^2/\lambda_S})$, and Case 3: $V_1(\sqrt{-m_h^2/\lambda_h},0)=V_1(0, \sqrt{-m_S^2/\lambda_S})$ for computing the order of phase transition and are depicted by different plotmarkers, i.e., circle, triangle and square, respectively. For case 1, $V_1^{\rm symm}(0,0) = V_1(\sqrt{-m_h^2/\lambda_h},0)$, the phase transition occurs directly from the symmetric to the broken phase with the varying Higgs field. The amalgamation of case 2 and case 3 gives two-step phase transition driven by the expectation value of the singlet field first, i.e., $(V_1^{\rm symm}(0,0) \rightarrow V_1(0, \sqrt{-m_S^2/\lambda_S}) \rightarrow V_1(\sqrt{-m_h^2/\lambda_h},0))$. Later, second transition occurs in the usual electroweak minimum which is dominated by the changing Higgs field and the EWBG occurs in the second transition which will be discussed in detail in \autoref{EWBG}. The first thing to be noted is that the contribution in the singlet direction only comes from the Higgs mass and the singlet mass. Another important thing is that the maximum allowed value for the interaction quartic coupling $\lambda_{h\Delta}$ for the Higgs field and the triplet field
from the Planck scale perturbativity is 1.25 and this will not contribute in the singlet direction. Therefore, the order of phase transition remains unaltered in the singlet direction (Case 2) with the variation in the triplet bare mass parameter (denoted by left and right triangle). For Case 1, the order of phase transition reduces with the increase in mass of the triplet and the order of phase transition is not strongly first-order for either of the masses and in contrast, it is strongly first-order for both the mass values for Case 3. There is no possibility for strongly first-order phase transition for Case 1, since the Higgs triplet interaction quartic coupling is restricted to $\lambda_{h\Delta}=1.25$ from Planck scale perturbativity because of the positive contribution from more number of degrees of freedom in comparison to \cite{Bandyopadhyay:2021ipw} (solely triplet $\lambda_{h\Delta}=1.95$). For Case 2, it happens because of the fact that the contribution in the singlet direction comes only from the Higgs mass and the singlet mass and this contribution is very less to achieve strongly first-order phase transition. In contrast to this, the strongly first-order phase transition is achieved for both 700 GeV and 1200 GeV for Case 3. Hence, the first transition from the symmetric phase to the singlet direction is not strongly first-order and later, the second transition from the singlet minimum to the usual EW minimum is strongly first order. The EWPT
 is sufficient to suppress the washout of the generated baryon number by sphaleron transitions.
 Hence, it gives the EWBG. 

\begin{figure}[h]
	\begin{center}
		\includegraphics[width=0.6\linewidth,angle=-0]{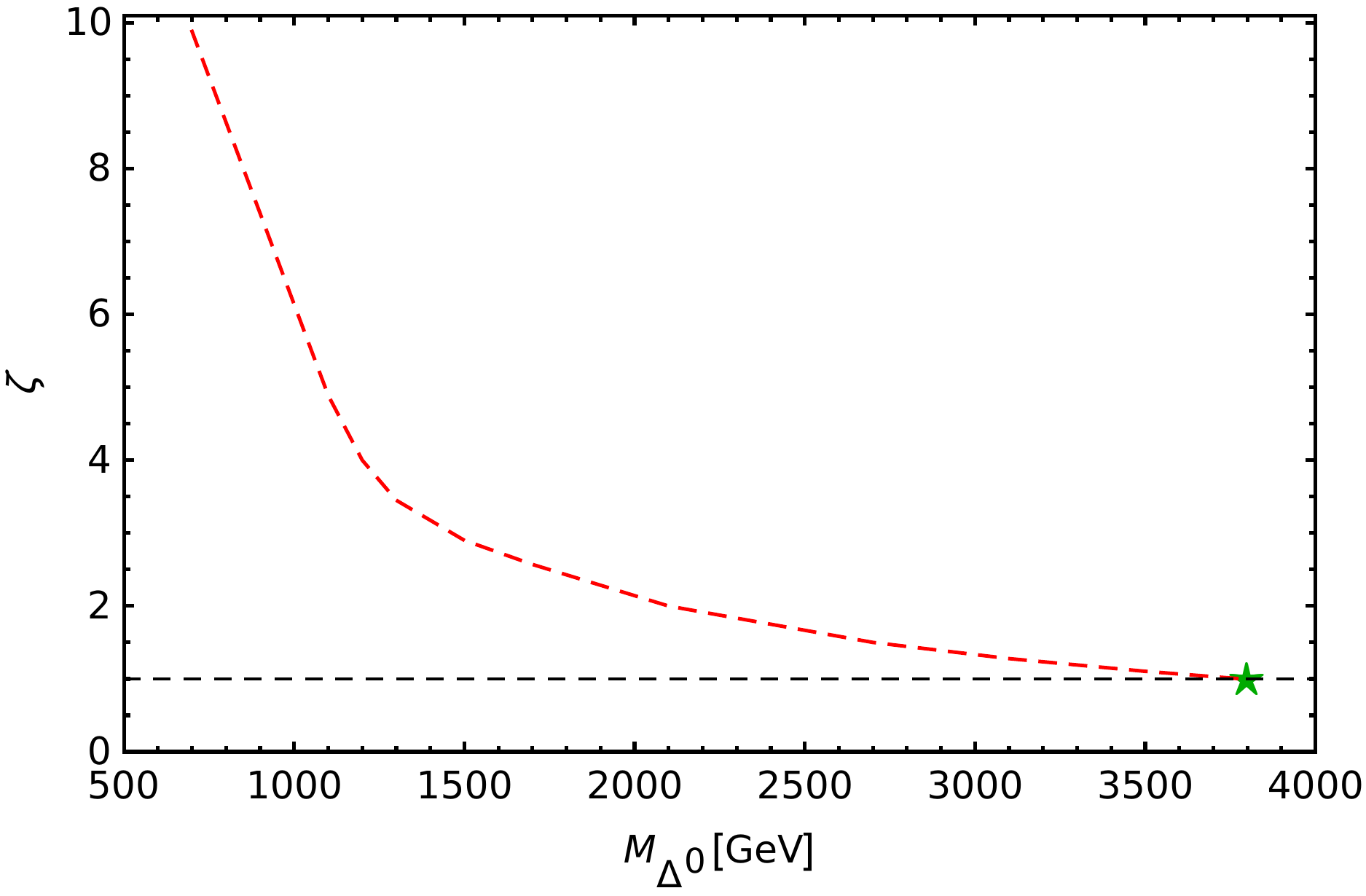}
		\caption{Variation of the order of phase transition $(\zeta=\frac{h_c}{T_c})$ with the triplet mass in GeV. The black dashed line corresponds to the strongly first-order criteria $\zeta=\frac{h_c}{T_c} \gsim 1$ and the green star denotes the upper mass bound on the triplet mass consistent with the measured Higgs boson mass and the Planck scale perturbative unitarity along with the first-order phase transition.
		}\label{fig3l}
	\end{center}
\end{figure}

The variation of the order of phase transition with the triplet mass in GeV is given in \autoref{fig3l}, denoted by the red dashed line. This plot is done for Case 3, where the transition occurs from the singlet vacuum to the usual electroweak vacuum. The black dashed line convey the strongly first-order phase transition criteria, i.e., $\zeta=\frac{h_c}{T_c} \gsim 1$.
The upper mass bound on the triplet gives the strongly first-order phase transition and it is consistent with the Planck scale stability, perturbativity, and the observed Higgs boson mass.
Here, its upper bound comes out to be 3.8 TeV, denoted by the green star. 
%
%
The other parameters are the same as those in BP2, which is consistent with the Planck scale perturbative unitarity and the measured Higgs boson mass in GeV.

The direct transition from the symmetric vacuum at high temperatures to the electroweak vacuum does not give strongly first-order phase transition for any of the triplet masses. However, the two-step transition
is driven by the changing Higgs field and strongly first-order.\\
In case of non-zero direct coupling of singlet with the triplet field ($\lambda_{\Delta s}$), the positive effect will be enhanced and the coupling relevant for the strength of electroweak phase transition ($\lambda_{\Delta h}$) would be restricted to lower values from Planck scale perturbativity. In case of very-weak limit ($\lambda_{\Delta s}=0.01-0.1$), the positive effect would not change the coupling, $\lambda_{\Delta h}$. For higher values of $\lambda_{\Delta s}$ $\sim$ 0.1-0.5, $\lambda_{\Delta h}$ reduces to 1.20, but this would still not have any significant change in the strength of phase transition, and the bound on the DM mass remains almost unaltered. Hence, it is absolutely fine to ignore the singlet-triplet coupling for the phase transition analysis.

The next section is devoted to the GW signatures generated during the first-order phase transition.
The expressions for computing the GW spectrum are given for completeness of the paper.

\section{Gravitational wave (GW) signatures}\label{gw}
The electroweak strongly first-order phase transition from the higher temperature symmetric phase to the broken phase occurs via nucleation of bubbles; the bubbles of the broken phase nucleate in the sea of symmetric phase, and ultimately the broken phase is achieved.
While expanding gives rise to GW,
the collision of these bubbles can be computed from three different contributions, i.e., Bubble wall collision \cite{Kosowsky,Turner,Huber_2008,Watkins,Marc,Caprini_2008}, Sound waves in the plasma \cite{Hindmarsh,Leitao:2012tx,Giblin:2013kea,Giblin:2014qia,Hindmarsh_2015} and the Magnetohydrodynamic turbulence in the plasma \cite{Chiara,Kahniashvili,Kahniashvili:2008pe,Kahniashvili:2009mf,Caprini:2009yp}. These contributions to the GW intensity are computed as follows \cite{Caprini:2015zlo};
\bea
h^2 \Omega_{GW}\simeq h^2 \Omega_{\phi}+h^2 \Omega_{sw}+h^2 \Omega_{turb}.
\eea
The first term from the envelope approximation \cite{Turner,Huber_2008,Watkins} via numerical simulations is given as;
\bea
h^2 \Omega_{env}(f) = 1.67 \times 10^{-5} \Big(\frac{\beta}{H}\Big)^{-2} \Big(\frac{\kappa_{\phi} \alpha}{1 + \alpha}\Big)^2 \Big(\frac{100}{g_*}\Big)^{1/3} \Big(\frac{0.11 v_w^3}{0.42 + v_w^2}\Big)\frac{3.8 (f/f_{env})^{2.8}}{1+2.8(f/f_{env})^{3.8}},
\eea
with 
\bea
\beta=\Big[HT\frac{d}{dT}\Big(\frac{S_3}{T}\Big)\Big]\Big|_{T_n}.
\eea
Here, $\beta$ defines the length of the time in which the phase transition completes, $T_n$ is the nucleation temperature at which the bubble nucleation starts, $H$ is the Hubble parameter.
$S_3$ is the Euclidean action of the background field that is computed for the critical bubble in the spherical polar coordinates as follows;
\bea
S_3= 4 \pi \int dr r^2 \Big[\frac{1}{2}(\partial_r \vec{\phi})^2+V_1(T)\Big]. 
\eea
Other important parameters for computing the GW background are $\alpha, \kappa_\phi, \kappa_v$, and $v_w$.
$\alpha$ describes the ratio of the energy density of vacuum to the radiation bath, which is being released during the phase transition defined as;
\bea
\alpha= \frac{\rho_{vac}}{\rho_{rad}^*},
\eea
where $\rho^*_{rad}=g_*\pi^2T_*^4/30$, $g_*$ being the number of relativistic degrees of freedom at temperature $T_*$ in plasma with $T_*=T_n$ in the absence of reheating.
Other additional parameters for assessing the GW frequencies are found as~\cite{Caprini:2015zlo, Shajiee:2018jdq,Kamionkowski:1993fg,Chao:2017vrq,Dev:2019njv,Paul};
\bea\label{alp}
\kappa_v & = & \frac{\rho_v}{\rho_{vac}}, \, \qquad  \kappa_\phi=\frac{\rho_\phi}{\rho_{vac}}= 1-\frac{\alpha_\infty}{\alpha}, \, \qquad v_w=\frac{1/\sqrt{3}+\sqrt{\alpha^2+2\alpha/3}}{1+\alpha}, \nn \\
\alpha_{\infty} & = & \frac{30}{24 \pi^2 g_*}\Big(\frac{v_n}{T_n}\Big)^2\Big[6\Big(\frac{m_W}{v}\Big)^2+3\Big(\frac{m_Z}{v}\Big)^2+6\Big(\frac{m_t}{v}\Big)^2\Big].
\eea 
$\kappa_v$ defines fraction of the vacuum energy that is being converted into the bulk motion of the fluid and $\kappa_\phi$ defines fraction of vacuum energy that is being converted into gradient energy of the Higgs-like field.
$v_w$ is the defined bubble wall velocity of the fluid, $v, v_n$ is the vev of the Higgs field at zero temperature and the nucleation temperature $T_n$, $m_W, m_Z$ and $m_t$ are the masses for the $W$ boson, $Z$ boson, and the top quark mass, respectively.
Finally, the expression for the peak frequency $f_{env}$, contributing to the GW intensity obtained from the bubble collisions, is given by
\bea
f_{\rm env}=16.5 \times 10^{-6} Hz \Big(\frac{0.62}{v^2_w-0.1 v_w+1.8}\Big)\Big(\frac{\beta}{H}\Big)\Big(\frac{T_n}{100 \rm GeV}\Big)\Big(\frac{g_*}{100}\Big)^{\frac{1}{6}}.
\eea

Secondly, the contribution from the sound waves in the plasma to the GW intensity is given as;
\bea
h^2 \Omega_{\rm SW}=2.65\times 10^{-6}\Big(\frac{\beta}{H}\Big)^{-1}v_w \Big(\frac{\kappa_v \alpha}{1+\alpha}\Big)^2 \Big(\frac{g_*}{100}\Big)^{-\frac{1}{3}}\Big(\frac{f}{f_{\rm SW}}\Big)^3\Big[\frac{7}{4+3\Big(\frac{f}{f_{\rm SW}}\Big)^2}\Big]^2,
\eea
where the parameter $\kappa_v$, given previously in \autoref{alp}, defining the fraction of latent heat that is converted to the bulk motion of the fluid, can now be rewritten as;
\bea \label{eq:4.22}
\kappa_v=\frac{\alpha_{\infty}}{\alpha}\Big[\frac{\alpha_{\infty}}{0.73+0.083 \sqrt{\alpha_{\infty}}+\alpha_{\infty}}\Big].
\eea
The peak frequency contribution of sound wave mechanisms $f_{\rm SW}$ to the GW spectrum produced is 
\bea
f_{\rm SW}=1.9\times 10^{-5}Hz \Big(\frac{1}{v_w}\Big)\Big(\frac{\beta}{H}\Big)\Big(\frac{T_n}{100 \rm GeV}\Big)\Big(\frac{g_*}{100}\Big)^{\frac{1}{6}}.
\eea
Lastly, the contribution from the Magnetohydrodynamic turbulence to the GW spectrum is given as \cite{Hogan:1986qda};
\bea
h^2 \Omega_{\rm turb}=3.35 \times 10^{-4}\Big(\frac{\beta}{H}\Big)^{-1}v_w \Big(\frac{\epsilon \kappa_v \alpha}{1+\alpha}\Big)^{\frac{3}{2}}\Big(\frac{g_*}{100}\Big)^{-\frac{1}{3}}\frac{(\frac{f}{f_{turb}})^3\Big(1+\frac{f}{f_{\rm turb}}\Big)^{-\frac{11}{3}}}{\Big(1+\frac{8\pi f}{h_*}\Big)},
\eea
where $\epsilon =0.1$ and $f_{\rm turb}$ is again the peak frequency contribution by the turbulence mechanism to the GW spectrum and is given as follows:
\bea
f_{\rm turb} = 2.7 \times 10^{-5} Hz \Big(\frac{1}{v_w}\Big)\Big(\frac{\beta}{H}\Big)\Big(\frac{T_n}{100 \rm GeV}\Big)\Big(\frac{g_*}{100}\Big)^{\frac{1}{6}}.
\eea
where,
\bea
h_*=16.5 \times 10^{-6} Hz \Big(\frac{T_n}{100 \rm GeV}\Big)\Big(\frac{g_*}{100}\Big)^{\frac{1}{6}}.
\eea
The upgraded expression for the $\kappa_v$ given in \autoref{eq:4.22} which is being used for this analysis is given as follows\cite{Ellis:2018mja, Espinosa:2010hh}:
\bea
\kappa_v \simeq \Big[\frac{\alpha_{\infty}}{0.73+0.083\sqrt{\alpha_{\infty}}+\alpha_{\infty}}\Big] .
\eea

\begin{figure}[H]
	\begin{center}
	\mbox{
		\subfigure[BP1]{\includegraphics[width=0.5\linewidth,angle=-0]{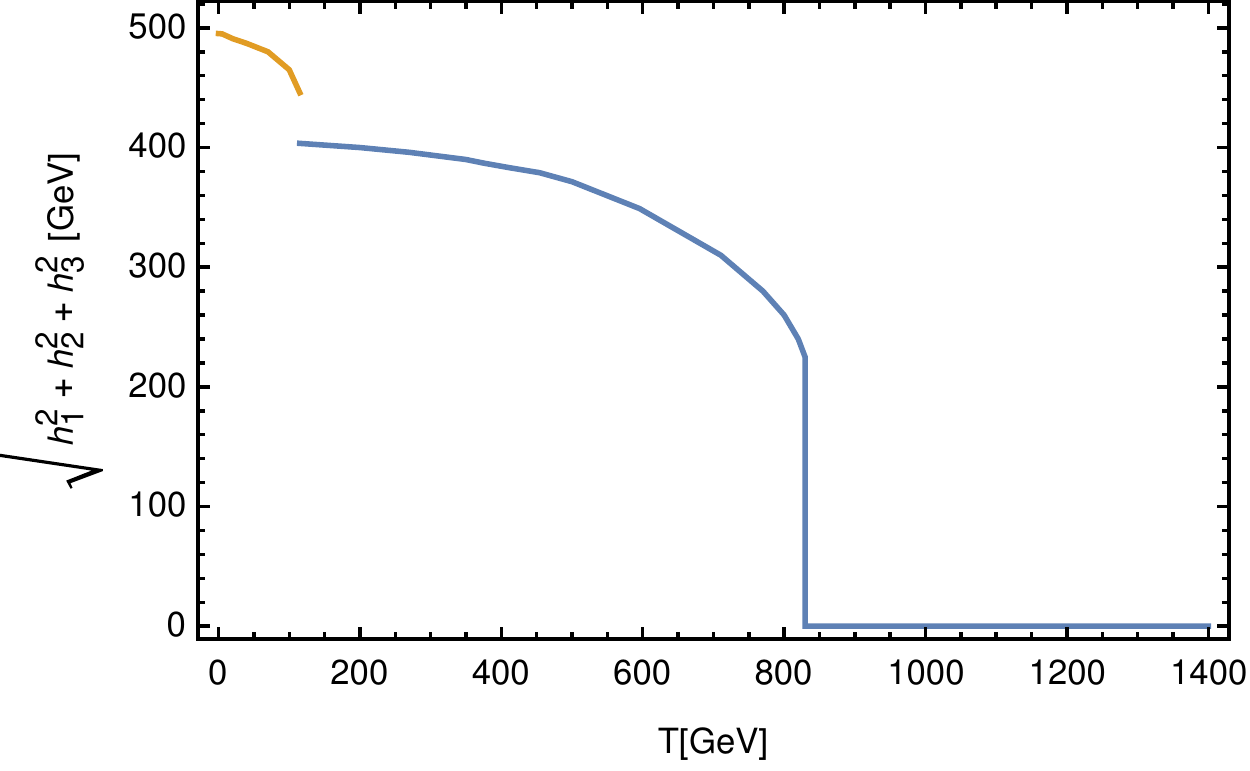}}
		\subfigure[BP2]{\includegraphics[width=0.5\linewidth,angle=-0]{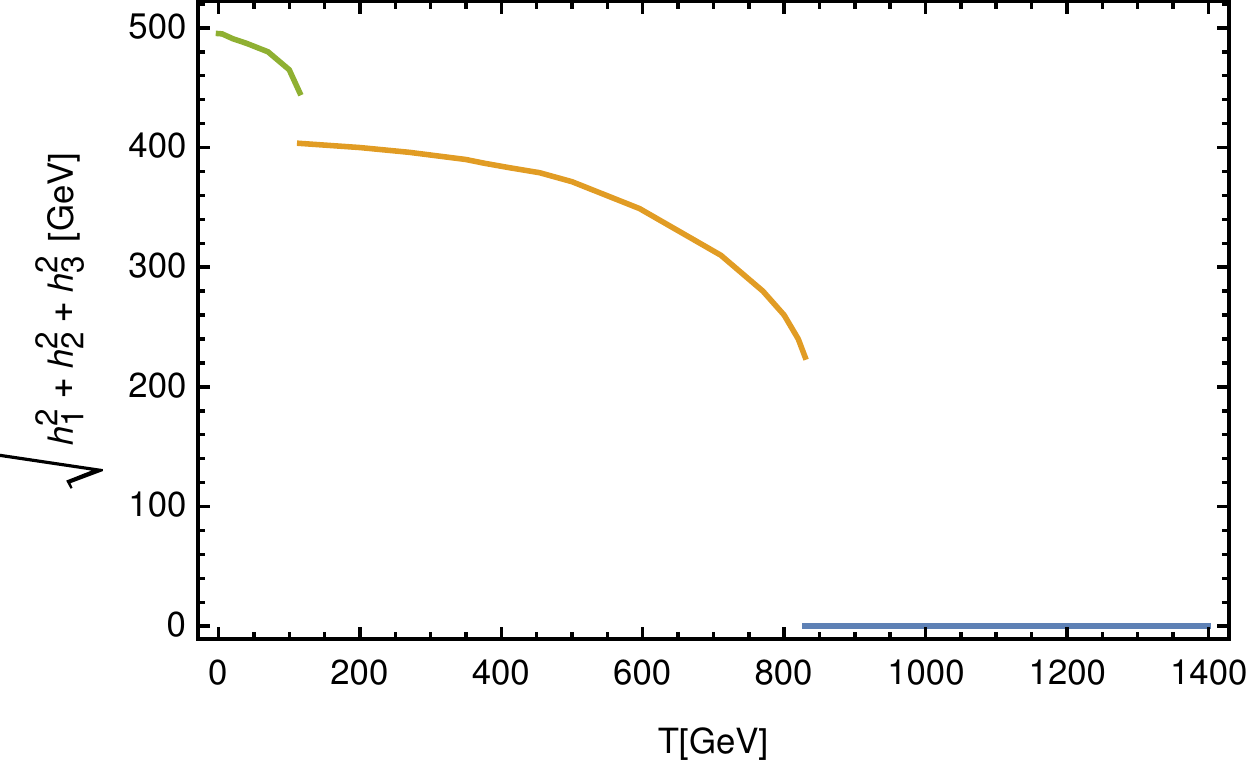}}}
		\caption{The variation of the minima of the scalar potential as a function of temperature in GeV, where $h_1, h_2$ and $h_3$ are the background fields for the SM Higgs doublet, inert triplet and the singlet, respectively. For BP1, one-step phase transition is acheived while BP2 proffers the two-step phase transition. 
		}\label{fig4l}
	\end{center}
\end{figure}

The effective potential in \autoref{eq:4.7} is implemented in the $\tt{CosmoTransition}$ \cite{Wainwright} package for computing the relevant parameters necessary for the computation of frequencies of the GWs.
The variation of the potential minima as a function of the temperature in GeV is given in \autoref{fig4l} for the benchmark points (BP1 and BP2) in \autoref{tab:table1} where, $h_1, h_2$ and $h_3$ are the background fields for the SM Higgs doublet, inert triplet and the singlet, respectively. For each benchmark point, we show the value of the critical temperature $T_c$, nucleation temperature $T_n$, "pattern" 1 or 2 indicates the one-step or two-step phase transition and "order" denotes the first and second-order phase transition. $\{h_1^{h,l}, h_2^{h,l}, h_3^{h,l}\}$ denotes the minima of the potential and the superscript $h$ and $l$ denotes the high-vev and the low-vev in the SM Higgs, triplet and the singlet direction, respectively, at a particular temperature. For BP1, there exists a one-step phase transition, which is first-order as given in \autoref{tab:table2}. At critical temperature $T_c=114.87$, the symmetry breaks in the SM Higgs and the singlet direction as $(h_1^l, h_2^l, h_3^l)=(189.23, 0.0, 403.1)$.
For BP2, the value of $\kappa$ is reduced to -0.019 and now, there exists two-step strongly first-order phase transition where, the symmetry first breaks in the singlet direction at $T_c=829.71$. The field values for this temperature are $(h_1^h, h_2^h, h_3^h)=(0.0, 0.0, 0.0)$ and $(h_1^l, h_2^l, h_3^l)=(0.0,0.0,230.89)$, which is second-order. Then, at $T_c=114.87$, the symmetry breaks in the SM Higgs and the singlet direction as $(h_1, h_2, h_3)=(189.23, 0.0, 403.1)$ similar to BP1. The nucleation temperature $T_n=112.33$ is same for both benchmark points. The other relevant parameters necessary for the GW spectrum are given below in \autoref{tab:table3} and the condition for the strongly first-order electroweak phase transition is $\zeta=\frac{h_c}{T_c} \geq 1$, where $h_c = \sqrt{(h_1^l-h_1^h)^2+(h_2^l-h_2^h)^2+(h_3^l-h_3^h)^2}$ at the critical temperature $T_c$. 


\begin{table}[h!]
	\begin{center}
		\begin{tabular}{|c|cccccc|}\hline
			& $i$ & pattern & $T_i$[GeV] & $\{h_1^h, h_2^h, h_3^h\}$ [GeV] & order & $\{h_1^l, h_2^l, h_3^l\}$ [GeV] \\ \hline
			BP1 & $T_c$ & 1 & 114.87 & $\{0.0, 0.0, 403.5\}$ & 1 & $\{189.23, 0.0, 403.1\}$ \\ 
			& $T_n$ & 1 & 112.33 & $\{0.0,0.0 , 403.5\}$ & 1 & $\{190.45, 0.0, 403.1\}$\\ \hline
			BP2 & $T_c$  & 2 & 829.71 & $\{0.0,0.0, 0.0\}$ & 2 & $\{0.0, 0.0, 230.89\}$\\ 
			& $T_c$  & 2 & 114.87 & $\{0.0, 0.0, 403.5\}$ & 1 & $\{189.23, 0.0, 403.1\}$ \\ 
			& $T_n$  & 2 & 112.33 & $\{0.0,0.0 , 403.5\}$ & 1 & $\{190.45, 0.0, 403.1\}$ \\ 
			\hline
		\end{tabular} 
	\end{center}
	\caption{Phase transition associated with the chosen benchmark points in \autoref{tab:table1}. For each benchmark point, we show the value of the critical temperature $T_c$, nucleation temperature $T_n$, "pattern" 1 or 2 indicates the one-step or two-step phase transition and "order" denotes the first and second-order phase transition. $\{h_1, h_2, h_3\}$ denotes the minima of the potential and the superscript $h$ and $l$ denotes the high-vev and the low-vev in the SM Higgs, triplet and the singlet direction, respectively, at a particular temperature.}	\label{tab:table2}
\end{table}

\begin{table}[h!]
	\begin{center}
		\begin{tabular}{|c|c|c|c|c|}\hline
			& $T_n$[GeV] & $\alpha$ & $\beta/H$ & $h_n/T_n$ \\ \hline
			BP1-2 & 112.33 & 0.436 & 290.68 & 1.69 \\ 
			\hline
		\end{tabular}
	\end{center}
	\caption{Thermal parameters needed for the frequency analysis of the SM+singlet+IT for the chosen BPs, where $T_n$ defines the nucleation temperature, $\alpha$ is defined as the strength of phase transition, $\beta$ defines the length of the time of phase transition and $h_n=\sqrt{(h_1^l-h_1^h)^2+(h_2^l-h_2^h)^2+(h_3^l-h_3^h)^2}$ at the nucleation temperature $T_n$. }	\label{tab:table3}
\end{table}

\begin{figure}[H]
	\begin{center}
			\includegraphics[width=0.5\linewidth,angle=-0]{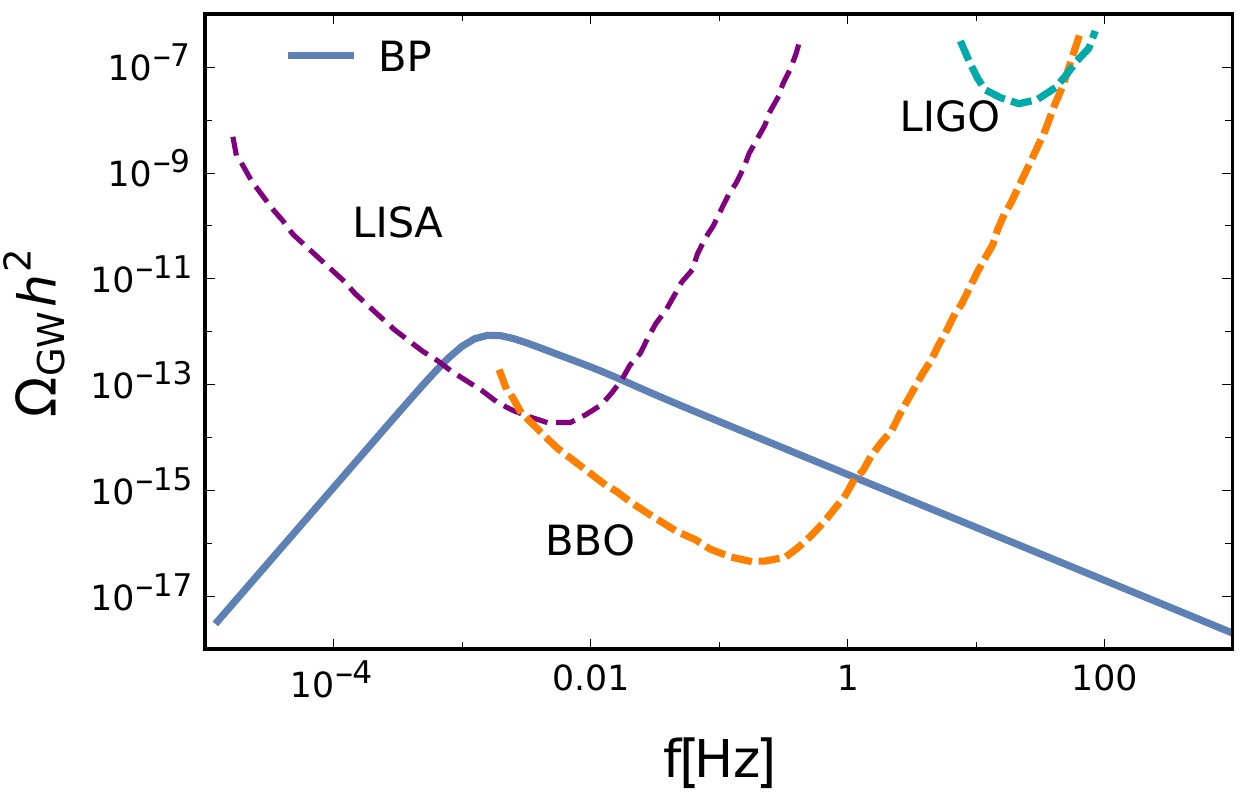}
		\caption{Variation of the GW intensity with the frequency in Hertz for the allowed BP which satisfies the measured Higgs boson mass bound, Planck scale stability and the perturbative unitarity, which is also consistent with the strongly first-order phase transition.
		}\label{fig5l}
	\end{center}
\end{figure}

Using the parameters given in \autoref{tab:table3}, the variation of the GW intensity with the frequency in Hertz is given in \autoref{fig5l}. The chosen BP is allowed from the measured SM Higgs boson mass bound, Planck scale stability and perturbative unitarity, which is also consistent with the strongly first-order phase transition. The nucleation temperature $T_n$ and the other relevant parameters are similar for both BP1 and BP2. Hence, the blue curve corresponds to the GW intensity variation for the chosen BPs. The purple, orange, and cyan colors respectively correspond to the GW spectrum for the detectable frequency range for LISA, LIGO, and BBO experiments. The frequency range between $6.978 \times 10^{-4} - 1.690 \times 10^{-2}$ Hz for the chosen BP lies in the detectable frequency range for LISA experiment.  It also lies in the detectable frequency range of BBO from $2.80\times 10^{-3}-1.096$. 
\\
The next section is devoted for the computation of the baryon asymmetry of the Universe arising from the additional phase introduced through the dimensional-5 effective operator. 

\section{Electroweak Baryogenesis (EWBG)}\label{EWBG}
\begin{figure}[H]
	\begin{center}
		\mbox{
		{\includegraphics[width=0.48\linewidth,angle=-0]{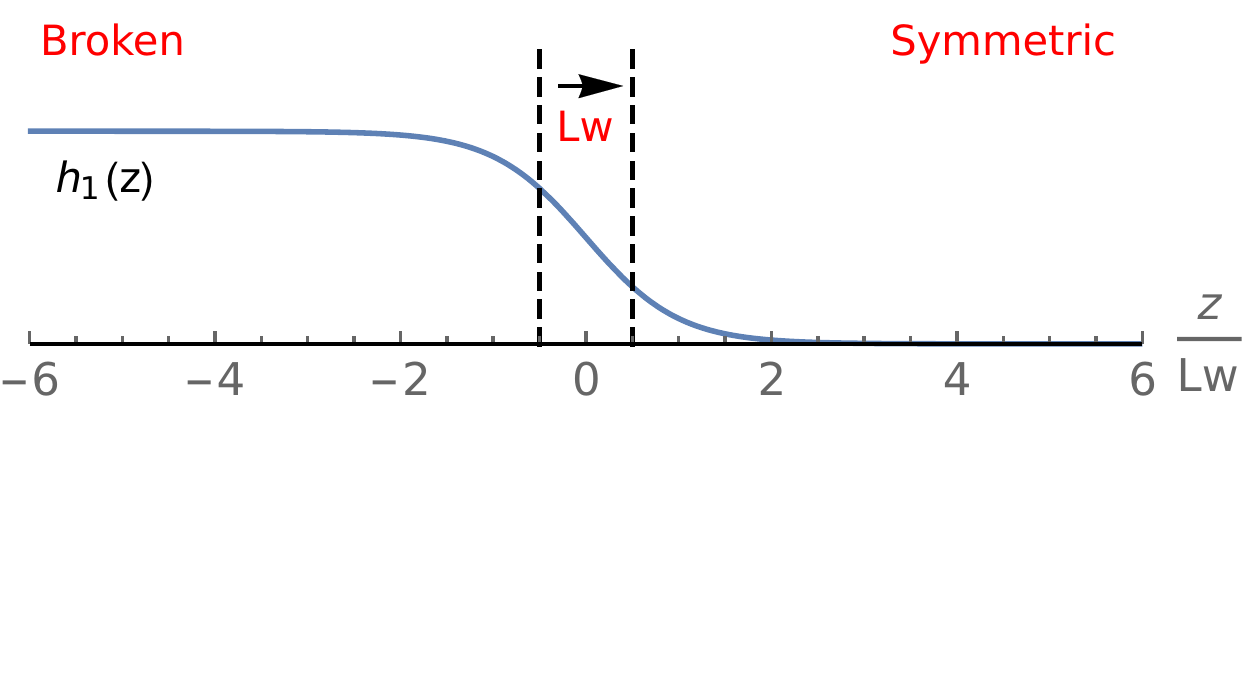}}
			{\includegraphics[width=0.48\linewidth,angle=-0]{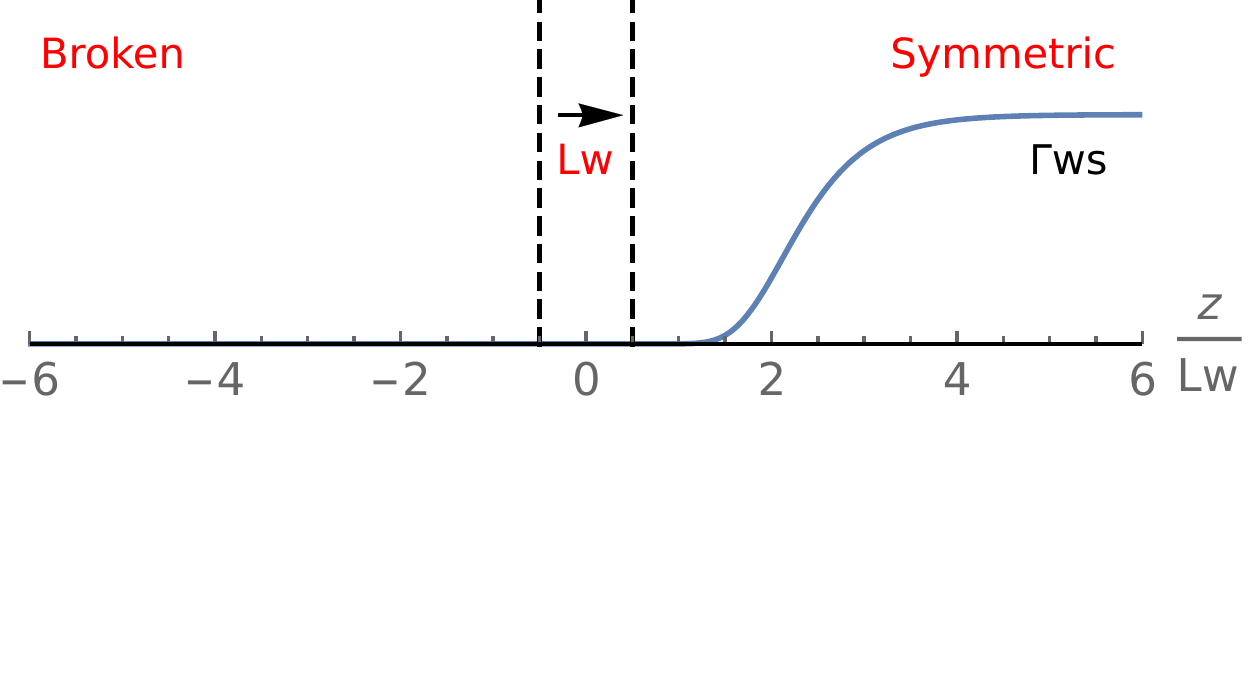}}}
		\caption{Profile of the Higgs field $h_1(z)$, and sphaleron transition rate $\Gamma_{ws}$ as a function of $\frac{z}{L_w}$. The Higgs vev becomes zero in the symmetric phase in front of the bubble wall and is non-zero inside the bubble. The weak sphaleron transition rate $\Gamma_{ws}$, is active in front of the bubble wall in the symmetric phase and suppressed in the broken phase.
		}\label{fig6l}
	\end{center}
\end{figure}

During the EWPT, the CP violation must be present within the bubble wall which separates the symmetric and broken phases for successful EWBG~\cite{Sakharov:1967dj}. Then this CP asymmetry is converted into the net baryon versus antibaryon excess in the symmetric phase in front of the bubble wall by the non-perturbative sphaleron processes . Later, when the bubble expands, this net baryon asymmetry diffuses inside the bubble wall, where these sphaleron transitions must be suppressed to avoid the washout of the net created baryon asymmetry. This suppression of sphaleron transitions brings another constraint that the EWPT must be strongly first-order ($v_c/T_c >1$)~\cite{Kuzmin:1985mm,PhysRevD.36.581}, where $v_c$ is the Higgs vev in the broken phase at the critical temperature $T_c$. Neither of these conditions are satisfied in case of SM, because the CP-violating phase provided by the CKM is too small, and hence the phase transition is a smooth crossover which puts lower mass bound on the Higgs mass from LEP \cite{Kajantie:1996mn}. As already mentioned, the EWPT in case of SM extended with singlet has been studied in detail \cite{Huber:2000mg,Espinosa:2011ax, PhysRevD.45.2685,Espinosa:1993bs,Profumo:2007wc,Ashoorioon:2009nf,Cline:2009sn}.
Here, we consider an additional CP-violating parameter, i.e., $\alpha$, in the dimension-5 operator, and we choose $\alpha = e^{i\pi/2}$ to maximize the CP-violation. Another important thing to note is that the tree-level barrier is crucial in variation of singlet vev during the EWPT. If the vev for the singlet is constant, then the scalar potential will have the similar shape as the SM potential at the tree-level, and there will be no tree-level barrier. Following the \autoref{eq:2.13}, the top quark mass can be written as $m_t=|m_t|e^{i \theta_t}$. 
At zero temperature,
it is always possible that $\theta_t$ can be absorbed by rotating the top quark field, {\it i.e.}, it is unphysical.
%
However, at finite temperature, the vev for singlet might change during EWPT, and the possibility of redefinition of top quark field goes away. Hence, the singlet vev must change during EWPT for a CP violation to give EWBG. This change in the vev is not at all guaranteed when the barrier is generated at loop-level, unlike tree-level. Hence, we assume that both $v$ and $x$ change along the $z$ direction which is perpendicular to the bubble wall, and the top mass is rewritten as:
\bea
m_t(z)= \frac{y_t}{\sqrt{2}}h_1(z)\Big(1+\frac{h_3(z)}{y_t \Lambda}\Big)=|m_t(z)|e^{i\theta_t(z)},\label{eq:7.1}
\eea
where $h_1(z)$ and $h_3(z)$ are respectively the field profiles around the bubble wall for the SM Higgs field and the singlet field $S$, $z$ being the coordinate perpendicular to the bubble wall.
With the assumption that the bubble wall is large enough, the bubble wall curvature is ignored, and it is assumed to be planar. Therefore, the field configurations for the SM Higgs and the singlet in the vicinity of the bubble wall are given as \cite{Bruggisser:2017lhc}:
\bea \label{eq:7.2}
h_1(z) &  = & \frac{v_{c} }{2}[1-\tanh(z/L_w)], \nn \\
h_3(z) & = & x_{c} + \frac{\Delta x_{c}}{2}[1-\tanh(z/L_w)],
\eea
where $L_w=\sqrt{\frac{(\Delta x_c^2 + v_c^2)}{8V_x}}$ is defined as the width of the bubble wall, $v_{c}$ is the vev of the Higgs-field at the critical temperature $T_c$, $\Delta x_{c}$ is the net change in the singlet vev at the critical temperature and $V_x$ is the height of the potential barrier at the critical temperature $T_c$ (the maximum height of the potential barrier along the path connecting the two minima). And the weak sphaleron transition rate is given by $\Gamma_{ws}=10^{-6} T e^{-a \frac{h_1(z)}{T}}$, where $a=37$. Using \autoref{eq:7.1} in \autoref{eq:7.2}, the complex phase for the varying top quark mass in \autoref{eq:7.1} is computed as; 
\bea \label{eq:7.3}
\theta_t(z)= \rm{Arc}\tan(\frac{\Delta \theta}{2}[1-tanh(z/L_w)]),
\eea
where, the CP phase is more or less described by $\tanh(z/	L_w)$, since the CP violation is usually small, $\Delta \theta << 1$.
The arbitary profile of the Higgs field $h_1(z)$ in \autoref{eq:7.2} is given in \autoref{fig6l}. The Higgs vev goes to zero in front of the bubble wall in the symmetric phase and is non-zero in the broken phase. The weak sphaleron transition rate $\Gamma_{ws}$ is active only in the symmetric phase where it converts the CP asymmetry into the net baryon number excess and inside the bubble, these transitions are strongly suppressed to avoid the washout of net baryon asymmetry created.

This space dependence lead to different dispersion relations for the particles and the anti-particles in the bubble wall, and this results in force terms in the transport equations from the WKB formalism. Thus, excess left-handed anti-quarks are obtained in front of the bubble wall, and this asymmetry is converted into net baryon asymmetry by the active sphaleron transitions in the symmetric phase. It is interesting to see that the non trivial phase $\theta(z)$ is sufficient enough to generate the desirable baryon asymmetry in this study. The energy scale in \autoref{eq:2.13} is chosen as $\Lambda=1000$ GeV for the further calculations.

For the top quark profile obtained from \autoref{eq:2.13}, we need to compute the chemical potentials for the particles involved in this particular interaction, i.e., for the left-handed $SU(2)$ doublet top $(\mu_{t,2})$, left-handed $SU(2)$ doublet bottom $(\mu_{b,2})$, left-handed $SU(2)$ singlet top quark $(\mu_{t^{c},2})$, and Higgs bosons $(\mu_{h,2})$, and also the plasma velocities corresponding to these. The net chemical potential for the left-handed quarks is given by;
\bea
\mu_{B_L}=\frac{1}{2}(1+4 \kappa_{t})\mu_{t,2}+\frac{1}{2}(1+4 \kappa_{b})\mu_{b,2} - 2 \kappa_{t}\mu_{t^{c},2},
\eea
where the $\kappa$ factors are basically the thermal averages. 
Now the net baryon asymmetry can be easily computed using the following formula;
\bea \label{eq:7.5}
\eta_B = \frac{n_B}{s}=\frac{405 \Gamma_{ws}}{4 \pi^2 v_w g_* T}\int_{0}^{\infty}dz \mu_{B_L}(z)e^{-\nu z},
\eea
where, $v_w$ is defined as the bubble wall velocity, $\Gamma_{ws}$ is the weak sphaleron rate, $\nu=\frac{45 \Gamma_{ws}}{4 v_w}$, $g_*\simeq106.75$ is the effective degrees of freedom in plasma, and the bubble wall velocity $v_w$ is chosen to be 0.1 for further calculations. The detailed computation for the chemical potential using the transport equations \cite{Bodeker:2004ws,Bruggisser:2017lhc} is given in \autoref{trans} .

\begin{figure}[H]
	\begin{center}
		\mbox{
			\subfigure[S(z)]{\includegraphics[width=0.48\linewidth,angle=-0]{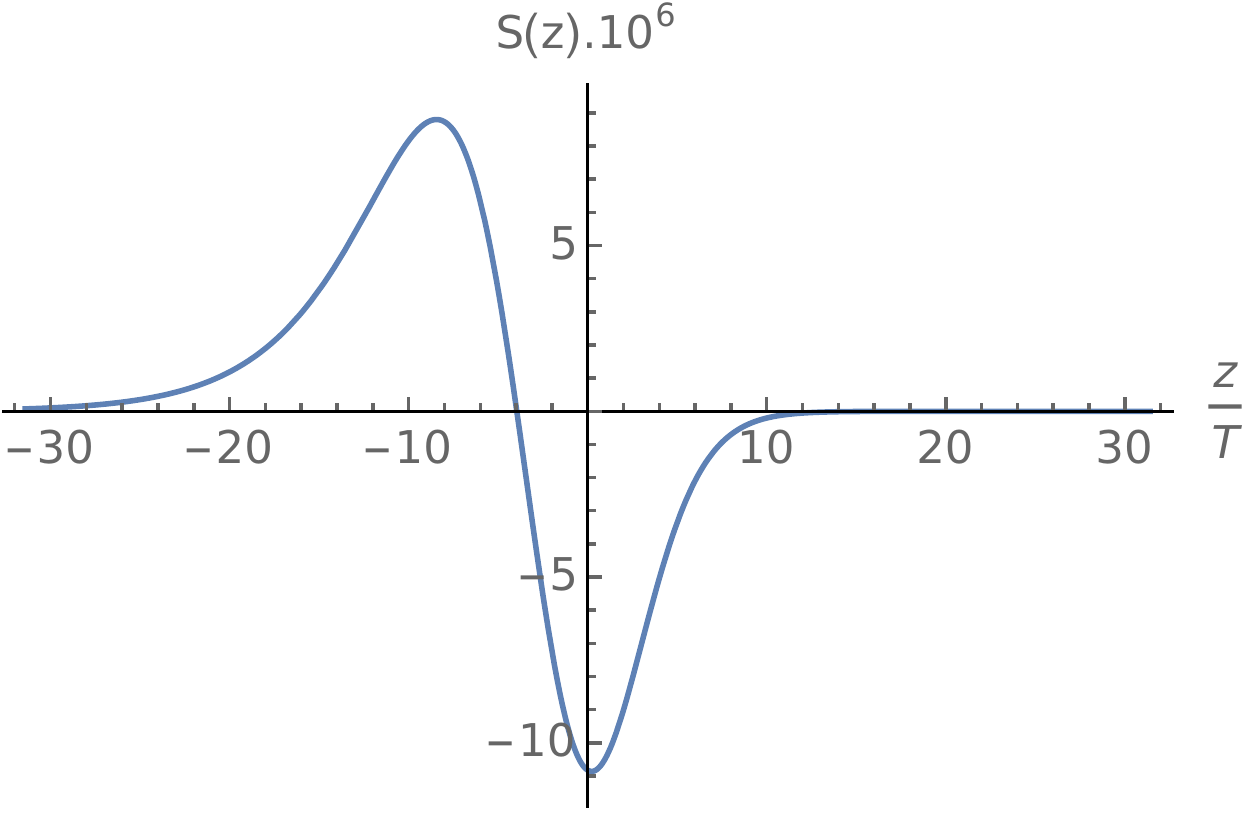}}
			\subfigure[$\mu_L(z)$]{\includegraphics[width=0.48\linewidth,angle=-0]{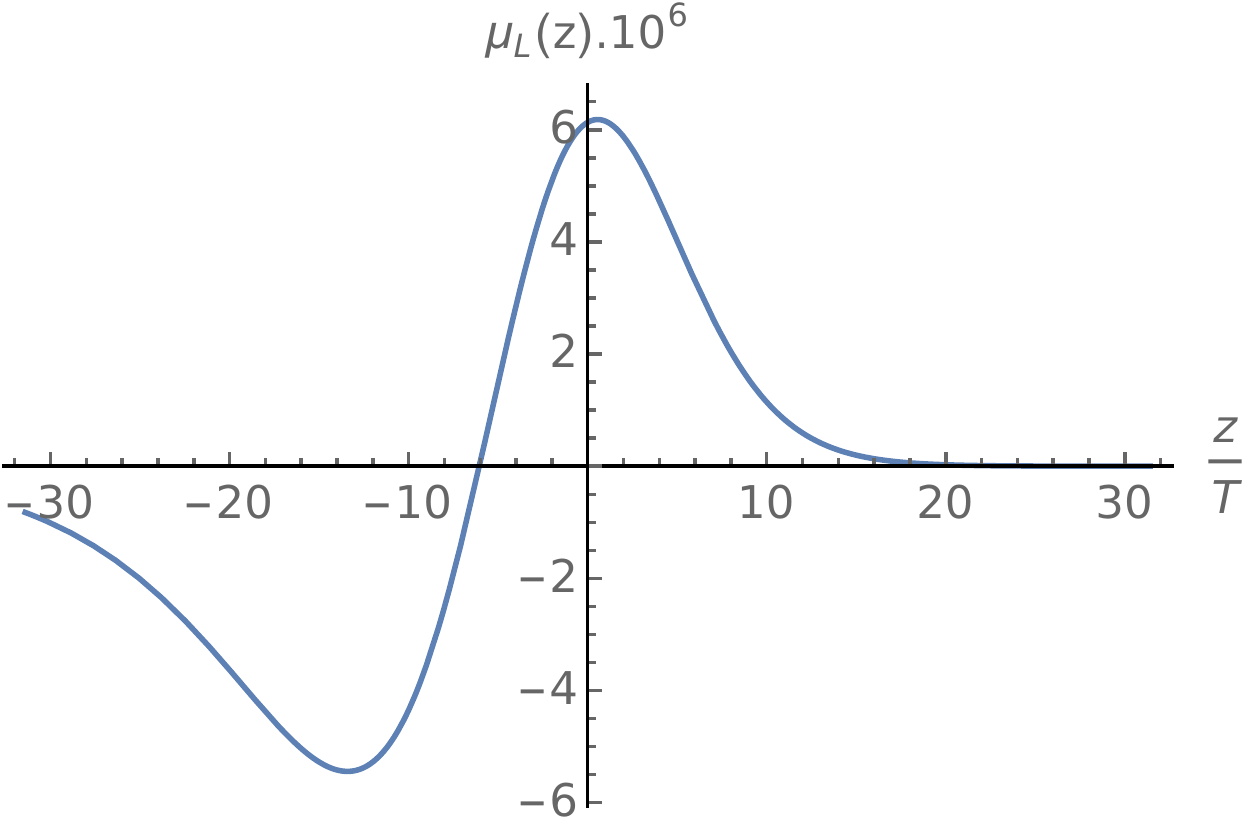}}}
		\caption{Variation of the source term $S(z)$ and the chemical potential $\mu_L(z)$ as a function of $\frac{z}{T}$. The source term on the right hand side of the diffusion equations, i.e., $K_{4,t}v_w m_t^2 \theta'' + K_{5,t}v_w(m_t^2)'\theta_t'$ for $L_w =\frac{8}{T}$, $v_w$=0.1 and $\Delta\theta=0.1$. The source term actually peaks within the bubble wall. 
		}\label{fig7l}
	\end{center}
\end{figure}

The variation of the source term, i.e., $K_{4,t}v_w m_t^2 \theta'' +K_{5,t}v_w(m_t^2)'\theta_t'$ with $\frac{z}{L_w}$ is given in \autoref{fig7l}(a). The $K_4$ and $K_5$ integrals are given in \autoref{trans}. The bubble wall width is approximately chosen from the tunnel bounce that was computed for the Froggatt-Nielsen models \cite{Baldes:2016gaf} and the values are typically between $\frac{3}{T}$ and $\frac{20}{T}$. In case of \autoref{fig7l}, this value is chosen to be $\frac{8}{T}$ and $\Delta\theta$ is fixed to 0.1. The $\frac{v_c}{T_c}$ factor which enters through the change in the top mass is 1.5. Using \autoref{eq:7.1} and \autoref{eq:7.3}, the source term can be easily computed and \autoref{fig7l}(a) depicts that the source term is typically peaked within the bubble wall. Then the chemical potential can be computed by solving the transport equations given in \autoref{trans} for estimating the net baryon asymmetry and the variation of the chemical potential as a function of $\frac{z}{T}$ is given in \autoref{fig7l}(b). Substituting this in \autoref{eq:7.5} the baryon asymmetry is computed which comes out to be $6.13 \times 10^{-11}$.

\section{Dark matter (DM) constraints}
After considering the theoretical constraints from the vacuum stability, perturbative unitarity, and EWPT, we focus on the DM constraints. Since the triplet field, being $Z_2$ odd, does not take part in the EWSB, the neutral scalar $M_\Delta^0$ serves as the DM candidate. The dominant annihilation mode which contributes to the DM relic density in case of ITM is \bl{$W^{\pm }W^{\mp}$}, discussed in ref.~\cite{Jangid:2020qgo} in details. The bound on the DM mass for the correct DM relic density from the WMAP \cite{WMAP:2012nax} and Planck experiments \cite{Planck:2013pxb} comes out to be 1.2 TeV where freeze-out scenario is assumed. This mass bound was not consistent with the strongly first-order phase transition in accordance with the measured Higgs boson mass and the Planck scale perturbativity in just the ITM case \cite{Bandyopadhyay:2021ipw}, and additional degrees of freedom were needed. 
Thanks to introduction of real singlet with vev in our model, the DM relic mass bound of 1.2 TeV can satisfy the strongly first-order phase transition along with the observed Higgs boson mass and the Planck scale perturbative unitarity.

\section{Electron electric dipole moment constraints (eEDM)}
The sufficient amount of CP-violation is crucial in explaining the observed Baryon asymmetry of the universe (BAU) during the electroweak phase transition. The recent bound on the electron EDM (eEDM) from ACME Collaboration \cite{ACME:2018yjb}, i.e., $d_e < 1.1 \times 10^{-29}$ e cm = $1.69 \times 10^{-16} $ $\rm GeV^{-1}$ impose stringent constraints on the beyond Standard Model scenarios which accomodates CP-violation. Since, CP-violation is so much crucial in explaining the BAU, it is important to check the consistency of the chosen parameter space with the eEDM constraints. The eEDM contribution comes only from the 2-loop Barr-Zee diagrams mediated by the top quark, and it is proportional to the imaginary part of $\alpha$ (CP-violating parameter), i.e, $d_e \propto Im\alpha$ as given in detail in  \cite{Keus:2017ioh}.
\bea\label{eq:9.1}
d_{e,t}^{2-loop} =\frac{e}{(3\pi^2)} \Big(\frac{\alpha G_F v}{\sqrt{2}\pi m_t}\Big)m_e \Big(\frac{v Im\alpha}{2\Lambda}\Big)\sin\theta \cos\theta \Big[-g(z_{th1}+g(z_{th2}))\Big],
\eea
where $g(z)$ is the loop function. \autoref{eq:9.1} predicts that for $m_{h}\sim m_{H}=125.5$ GeV, and $g(z_{th1}\sim g(z_{th2}))$, the two-loop contribution to $d_e$ is reduced. As $m_{H}$ approaches $m_h$, a large region of the parameter space will survive the eEDM bounds as shown in Figure 5 in \cite{Keus:2017ioh}. The benchmark point chosen in our considered scenario for the EWPT and explaining the BAU gives $m_H$ as 177.26 GeV which is consistent with the eEDM bound.\\

\section{Conclusion}\label{conc}
We have analyzed a SM extended with a real singlet and a $Y=0$ Higgs triplet considering an additional dimension-5 term. The parameter points are chosen to be consistent with the Planck scale stability and the perturbative unitarity using two-loop $\beta-$ functions. The interaction quartic coupling between Higgs and the triplet $\lambda_{h\Delta}$, which is crucial for determining the triplet mass, is restricted to 1.25 from Planck scale perturbativity in presence of an additional singlet field. While the bound has to be 1.95 in case of inert triplet solely. 
We have also studied that the presence of singlet helps in achieving the two-step phase transition where the symmetry firstly breaks in the singlet field direction, and later,  to the usual electroweak minima. The phase transition from the singlet field to the electroweak minima is found to be strongly first-order consistent with the measured Higgs boson mass of 125.5 GeV.

We have found that the lightest stable neutral component from the Higgs triplet can be a DM candidate that satisfy the observed relic density for 1.2 TeV of the DM mass where the DM mass below 1.2 TeV is not allowed because of
under-abundance.
We have also checked that our DM is consistent with the strongly first-order phase transition along with the observed Higgs boson mass and the Planck scale perturbativity.
%
If we introduce the inert triplet only,
the upper bound on the DM mass to satisfy the strongly first-order phase transition is 259.0 GeV that would already be ruled out by current experiments such as LHC.
Thanks to the singlet boson that causes the possibility of a two-step phase transition, however, the bound shifts to 3.8 TeV that is still within the range of allowed space.

Then, we have studied electroweak baryogenesis using the CP asymmetry generated by the fermionic sector of the dimension-5 term. 
The chosen benchmark point satisfies all possible constraints; {\it i.e.}, strongly first-order phase transition, Planck scale perturbativity, and measured Higgs boson mass.
Moreover, the DM constraint from relic density also satisfies the correct baryon to entropy ratio $(6.13 \times 10^{-11})$.

We have also explored the GW and the corresponding frequency lies in the detectable frequency range of the LISA ($6.978 \times 10^{-4} - 1.690 \times 10^{-2}$) Hz and BBO experiments.

\section{Acknowledgement}
 This research was supported by an appointment to the JRG Program at the APCTP through the Science and Technology Promotion Fund and Lottery Fund of the Korean Government. This was also supported by the Korean Local Governments - Gyeongsangbuk-do Province and Pohang city (H.O.).
This research was also supported by an appointment to the YST Program at the APCTP through the Science and Technology Promotion Fund and Lottery Fund of the Korean Government. This was also supported by the Korean Local Governments - Gyeongsangbuk-do Province and Pohang city (S.J.). S.J. thanks Thomas Konstandin and Prof. Ligong Bian for useful guidance in computing the baryon asymmetry. SJ also thanks Anirban Karan for useful discussions.
\appendix

\section{Two-loop $\beta$-functions for dimensionless couplings} \label{betaf1}
\subsection{Scalar Quartic Couplings}\label{A1}
\footnotesize{
	\begingroup
	\allowdisplaybreaks
	\begin{align*}
		\beta_{\lambda_h} \ =  \ &
		\frac{1}{16\pi^2} \Bigg[+\frac{27}{200} g_{1}^{4} +\frac{9}{20} g_{1}^{2} g_{2}^{2} +\frac{9}{8} g_{2}^{4} -\frac{9}{5} g_{1}^{2} \lambda -9 g_{2}^{2} \lambda +24 \lambda_h^{2} +2 \lambda_{hs}^{2} +9 \lambda_{h\Delta}^{2} +12 \lambda_h \mbox{Tr}\Big({Y_d  Y_{d}^{\dagger}}\Big) \nonumber \\ 
		&\qquad +4 \lambda_h \mbox{Tr}\Big({Y_e  Y_{e}^{\dagger}}\Big) +12 \lambda_h \mbox{Tr}\Big({Y_u  Y_{u}^{\dagger}}\Big) -6 \mbox{Tr}\Big({Y_d  Y_{d}^{\dagger}  Y_d  Y_{d}^{\dagger}}\Big) -2 \mbox{Tr}\Big({Y_e  Y_{e}^{\dagger}  Y_e  Y_{e}^{\dagger}}\Big) -6 \mbox{Tr}\Big({Y_u  Y_{u}^{\dagger}  Y_u  Y_{u}^{\dagger}}\Big) \Bigg] \nonumber \\
		& +\frac{1}{(16\pi^2)^2}\Bigg[-\frac{3411}{2000} g_{1}^{6} -\frac{1677}{400} g_{1}^{4} g_{2}^{2} -\frac{317}{80} g_{1}^{2} g_{2}^{4} +\frac{277}{16} g_{2}^{6} +\frac{1887}{200} g_{1}^{4} \lambda_h +\frac{117}{20} g_{1}^{2} g_{2}^{2} \lambda_h -\frac{29}{8} g_{2}^{4} \lambda_h +\frac{108}{5} g_{1}^{2} \lambda_h^{2} +108 g_{2}^{2} \lambda_h^{2} \nonumber \\ 
		&\qquad -312 \lambda_h^{3} -20 \lambda_h \lambda_{hs}^{2} -16 \lambda_{hs}^{3} +\frac{105}{4} g_{2}^{4} \lambda_{h\Delta} +\frac{171}{2} g_{2}^{2} \lambda_{h\Delta}^{2} -90 \lambda_h \lambda_{h\Delta}^{2} -132 \lambda_{h\Delta}^{3} \nonumber \\ 
		&\qquad +\frac{1}{20} \Big(-5 \Big(64 \lambda_h \Big(-5 g_{3}^{2}  + 9 \lambda_h \Big) -90 g_{2}^{2} \lambda_h  + 9 g_{2}^{4} \Big) + 9 g_{1}^{4}  + g_{1}^{2} \Big(50 \lambda_h  + 54 g_{2}^{2} \Big)\Big)\mbox{Tr}\Big({Y_d  Y_{d}^{\dagger}}\Big) \nonumber \\ 
		&\qquad -\frac{3}{20} \Big(15 g_{1}^{4}  -2 g_{1}^{2} \Big(11 g_{2}^{2}  + 25 \lambda_h \Big) + 5 \Big(-10 g_{2}^{2} \lambda_h  + 64 \lambda_h^{2}  + g_{2}^{4}\Big)\Big)\mbox{Tr}\Big({Y_e  Y_{e}^{\dagger}}\Big) -\frac{171}{100} g_{1}^{4} \mbox{Tr}\Big({Y_u  Y_{u}^{\dagger}}\Big) \nonumber \\ 
		&\qquad +\frac{63}{10} g_{1}^{2} g_{2}^{2} \mbox{Tr}\Big({Y_u  Y_{u}^{\dagger}}\Big) -\frac{9}{4} g_{2}^{4} \mbox{Tr}\Big({Y_u  Y_{u}^{\dagger}}\Big) +\frac{17}{2} g_{1}^{2} \lambda_h \mbox{Tr}\Big({Y_u  Y_{u}^{\dagger}}\Big) +\frac{45}{2} g_{2}^{2} \lambda_h \mbox{Tr}\Big({Y_u  Y_{u}^{\dagger}}\Big) \nonumber \\ 
		&\qquad +80 g_{3}^{2} \lambda_h \mbox{Tr}\Big({Y_u  Y_{u}^{\dagger}}\Big) -144 \lambda_h^{2} \mbox{Tr}\Big({Y_u  Y_{u}^{\dagger}}\Big) +\frac{4}{5} g_{1}^{2} \mbox{Tr}\Big({Y_d  Y_{d}^{\dagger}  Y_d  Y_{d}^{\dagger}}\Big) -32 g_{3}^{2} \mbox{Tr}\Big({Y_d  Y_{d}^{\dagger}  Y_d  Y_{d}^{\dagger}}\Big) \nonumber \\ 
		&\qquad -3 \lambda_h \mbox{Tr}\Big({Y_d  Y_{d}^{\dagger}  Y_d  Y_{d}^{\dagger}}\Big) -42 \lambda_h \mbox{Tr}\Big({Y_d  Y_{u}^{\dagger}  Y_u  Y_{d}^{\dagger}}\Big) -\frac{12}{5} g_{1}^{2} \mbox{Tr}\Big({Y_e  Y_{e}^{\dagger}  Y_e  Y_{e}^{\dagger}}\Big) - \lambda_h \mbox{Tr}\Big({Y_e  Y_{e}^{\dagger}  Y_e  Y_{e}^{\dagger}}\Big) \nonumber \\ 
		&\qquad -\frac{8}{5} g_{1}^{2} \mbox{Tr}\Big({Y_u  Y_{u}^{\dagger}  Y_u  Y_{u}^{\dagger}}\Big) -32 g_{3}^{2} \mbox{Tr}\Big({Y_u  Y_{u}^{\dagger}  Y_u  Y_{u}^{\dagger}}\Big) -3 \lambda \mbox{Tr}\Big({Y_u  Y_{u}^{\dagger}  Y_u  Y_{u}^{\dagger}}\Big) +30 \mbox{Tr}\Big({Y_d  Y_{d}^{\dagger}  Y_d  Y_{d}^{\dagger}  Y_d  Y_{d}^{\dagger}}\Big) \nonumber \\ 
		&\qquad -12 \mbox{Tr}\Big({Y_d  Y_{d}^{\dagger}  Y_d  Y_{u}^{\dagger}  Y_u  Y_{d}^{\dagger}}\Big) +6 \mbox{Tr}\Big({Y_d  Y_{u}^{\dagger}  Y_u  Y_{d}^{\dagger}  Y_d  Y_{d}^{\dagger}}\Big) -6 \mbox{Tr}\Big({Y_d  Y_{u}^{\dagger}  Y_u  Y_{u}^{\dagger}  Y_u  Y_{d}^{\dagger}}\Big) \nonumber \\ 
		&\qquad +10 \mbox{Tr}\Big({Y_e  Y_{e}^{\dagger}  Y_e  Y_{e}^{\dagger}  Y_e  Y_{e}^{\dagger}}\Big) +30 \mbox{Tr}\Big({Y_u  Y_{u}^{\dagger}  Y_u  Y_{u}^{\dagger}  Y_u  Y_{u}^{\dagger}}\Big) \Bigg] \, . \\
	\end{align*}
	\subsection{Gauge Couplings}
	\footnotesize{
		\begingroup
		\allowdisplaybreaks
		\begin{align*}
			\beta_{g_1} \  = \ &  
			\frac{1}{16\pi^2}\Bigg[\frac{41}{10} g_{1}^{3}\Bigg]+\frac{1}{(16\pi^2)^2}\Bigg[ \frac{1}{50} g_{1}^{3} \Big(135 g_{2}^{2}  + 199 g_{1}^{2}  -25 \mbox{Tr}\Big({Y_d  Y_{d}^{\dagger}}\Big)  + 440 g_{3}^{2}  -75 \mbox{Tr}\Big({Y_e  Y_{e}^{\dagger}}\Big)  -85 \mbox{Tr}\Big({Y_u  Y_{u}^{\dagger}}\Big) \Big)\Bigg] \, .  \\
			\beta_{g_2} \  = \ &  
			\frac{1}{16\pi^2}\Bigg[-\frac{17}{6} g_{2}^{3}\Bigg]+\frac{1}{(16\pi^2)^2}\Bigg[ \frac{1}{30} g_{2}^{3} \Big(-15 \mbox{Tr}\Big({Y_e  Y_{e}^{\dagger}}\Big)  + 27 g_{1}^{2}  + 360 g_{3}^{2}  + 455 g_{2}^{2}  -45 \mbox{Tr}\Big({Y_d  Y_{d}^{\dagger}}\Big)  -45 \mbox{Tr}\Big({Y_u  Y_{u}^{\dagger}}\Big) \Big)\Bigg] \, .  \\
			\beta_{g_3} \  = \ &  
			\frac{1}{16\pi^2}\Bigg[-7 g_{3}^{3}\Bigg]+\frac{1}{(16\pi^2)^2}\Bigg[-\frac{1}{10} g_{3}^{3} \Big(-11 g_{1}^{2}  + 20 \mbox{Tr}\Big({Y_d  Y_{d}^{\dagger}}\Big)  + 20 \mbox{Tr}\Big({Y_u  Y_{u}^{\dagger}}\Big)  + 260 g_{3}^{2}  -45 g_{2}^{2} \Big)\Bigg] \, .  \\
		\end{align*}
		\endgroup
 
\section{Dimensionally reduced parameters}\label{3dexp}
The scalar potential given in \autoref{eq:2.6} for the SM extended with a singlet and an inert Higgs triplet scenario in the dimensionally reduced 3D effective theories (DR3EFTs) is given as:
\bea
V_0^{\rm eff} & = & m_{h,3}^2  \Phi^\dagger \Phi+m_{\Delta,3}^2  Tr(\Delta^\dagger \Delta)+\lambda_{h,3}|\Phi^\dagger \Phi|^2+\lambda_{\Delta,3}(Tr|\Delta^\dagger \Delta|)^2+\lambda_{h\Delta,3}\Phi^\dagger \Phi Tr(\Delta^\dagger \Delta) +m_{S,3}^2  S^2 + \lambda_{S,3}S^4 
+ \kappa_3 {S}^3 \nn \\
 && + \lambda_{hs,3}(\Phi^\dagger\Phi)(S^2). 
\label{eq:8.1}
\eea
	The matching relations for the quartic couplings and the bare masses (which are the tree-level parameters) are computed as follows \cite{Schicho:2021gca,Niemi:2021qvp,Niemi:2018asa}:
	\bea
	\lambda_{h,3} & = & T\Big[\lambda_h (\Lambda) + \frac{1}{(4\pi)^2}\Big(\frac{2-3L_b}{16}(3g_2^4+2g_2^2g_1^2+g_1^4)+ N_c L_f(y_t^4-2\lambda_hy_t^2)+L_b\Big(\frac{3}{2}(3g_2^2+g_1^2)\lambda_h-12\lambda_h^2-\frac{1}{2} \lambda_{hs}\Big)\Big) \nn \\
	&& + \frac{1}{(4\pi)^2}\Big[\frac{1}{8}\Big(3g_2^4 +g_1^4 + 2 g_2^2 g_1^2 \Big)+3 L_f\Big(y_t^4 - 2 \lambda_1 y_t^2\Big)-L_b \Big(\frac{3}{16}\Big(3 g_2^4+g_1^4+2g_1^2 g_2^2\Big)
	 -\frac{3}{2}\Big(3g_2^2 + g_1^2 - 8 \lambda_h\Big)\lambda_h + \frac{3}{4}(2\lambda_{ht})^2\Big)\Big] \nn ,\\
	\lambda_{s,3} & = & \frac{T}{4}\Big[4\lambda_s(\Lambda)-\frac{1}{(4\pi)^2}L_b((2\lambda_{hs})^2+9(4 \lambda_s)^2)\Big], \nn\\
	\lambda_{hs,3} & = & \frac{T}{2}\Big[2\lambda_{hs}(\Lambda)+ \frac{2\lambda_{hs}}{(4\pi)^2}\Big(L_b\Big(\frac{3}{4}(3g_2^2+g_1^2)-6\lambda_1-4\lambda_{hs}-12\lambda_s\Big)-N_cL_f y_t^2)\Big] \nn , \\
	\lambda_{\Delta,3} & = & \frac{T}{4}\Big[4\lambda_\Delta(\Lambda) + \frac{1}{(4\pi)^2} \Big[4 g_2^4 - L_b((2\lambda_{h\Delta})^2 + 11 (4\lambda_\Delta)^2 -48 g_2^2 \lambda_\Delta + 6 g_2^4)\Big]\Big], \nn\\
    \lambda_{h\Delta,3} & = & \frac{T}{2} \Big[2\lambda_{h\Delta}(\Lambda) + \frac{1}{(4\pi)^2}\Big[2g_2^4-6\lambda_{h\Delta}y_t^2 L_f -L_b\Big(8 \lambda_{h\Delta}^2 + 40 \lambda_{h\Delta}\lambda_\Delta + 3 g_2^4 + 12 \lambda_{h\Delta}\lambda_h -\frac{3}{2}\lambda_{h\Delta}(g_1^2 + 11 g_2^2)\Big)\Big]\Big] \nn , \\
    \kappa_3 & = & \frac{\sqrt{T}}{3}\Big[3 \kappa(\Lambda)-\frac{3L_b}{(4\pi)^2}\Big(36 \lambda_s \kappa \Big)\Big].
	\eea
	where 
	\bea
	L_b & = & \ln\Big(\frac{\Lambda^2}{T^2}\Big)-2[\ln(4\pi)-\gamma],\\
	L_f & = & L_b + 4 \ln2.
	\eea
	Here, $L_b$ and $L_f$ are logarithms that arise frequently from one-loop bosonic and fermionic sum integrals with $\Lambda$ is the $\overline{MS}$ scale and $\gamma$ is the Euler-Mascheroni constant. The expressions for the two-loop mass parameters are computed as follows:
	\bea
	m_{h,3}^2 & = & (m_{h,3}^2)_{\rm SM} + \frac{T^2}{12}\lambda_{hs}(\Lambda)-\frac{L_b}{(4\pi)^2}\Big(2\lambda_{hs}m_S^2(\Lambda)\Big)+\frac{1}{(4\pi)^2}\Big(\frac{3}{4}(3g_2^2+g_1^2)L_b-N_c y_t^2L_f\Big)\Big(\frac{T^2}{12}\lambda_{hs}\Big)+\frac{1}{(4\pi)^4}\Big[9(3+2L_b+L_b^2)(\lambda_{hs}\kappa^2)\Big] \nn \\
	&&-\frac{2T^2}{(4\pi)^2}L_b\lambda_{hs}\Big(\frac{1}{4}\lambda_h + \frac{5}{12}\lambda_{hs}
	 +\frac{1}{2}\lambda_s\Big)-\frac{2}{(4\pi)^2}\lambda_{hs,3}^2\Big(c+ \ln\Big(\frac{3T}{\Lambda_{3d}}\Big)\Big) + \frac{T^2}{4}\lambda_{h\Delta}(\Lambda) + \frac{1}{16 \pi^2}\Big[+6\lambda_{h\Delta} m_\Delta^2L_b + T^2 \Big(\frac{5}{24}g_2^4 +\lambda_{h\Delta}g_2^2  \nn \\
	 && - \frac{3}{4}\lambda_{h\Delta}y_t^2 L_f + L_b \Big(-\frac{7}{16}g_2^4 - \frac{5}{2}\lambda_{h\Delta}^2- \lambda_{h\Delta}\lambda_\Delta + \frac{33}{16}\lambda_{h\Delta}g_2^2 + \frac{3}{16}\lambda_{h\Delta}g_1^2 - \frac{3}{2}\lambda_{h\Delta}\lambda_h\Big)+\Big(c+ \ln(\frac{3T}{\Lambda_{3d}})\Big)\Big(-6\lambda_{h\Delta,3}^2 \nn \\
	 && +12\lambda_{h\Delta,3}g_{2,3}^2- \frac{3}{4}g_{2,3}^4\Big)\Big)\Big],
	\eea
	where
	\bea
	(\mu_3^2)_{\rm SM} & = & \mu^2(\Lambda) + \frac{T^2}{12}\Big(\frac{3}{4}(3g_2^2(\Lambda)+g_1^2(\Lambda))+N_cy_t^2(\Lambda)+6\lambda_1(\Lambda)\Big)+ \frac{\mu^2(\Lambda)}{(4\pi)^2}\Big(\Big(\frac{3}{4}(3g_2^2+g_1^2)-6\lambda_1\Big)L_b-N_cy_t^2L_f\Big) \nn \\
	&& + \frac{T^2}{(4\pi)^2}\Big[\frac{167}{96}g_2^4 + \frac{1}{288}g_1^4 -\frac{3}{16}g_2^2g_1^2 + \frac{(1+3L_b)}{4}\lambda_1(3g_2^2+g_1^2)+L_b\Big(\frac{17}{16}g_2^4 - \frac{5}{48}g_1^4-\frac{3}{16}g_2^2g_1^2-6\lambda_1^2\Big) \nn \\
	&& +\frac{1}{T^2}\Big(c+ \ln \Big(\frac{3T}{\Lambda_{3d}}\Big)\Big)\Big(\frac{39}{16}g_{2,3}^4+ 12 g_{2,3}^2h_3 - 6 h_3^2 + 9 g_{2,3}^2\lambda_{1,3}-12\lambda_{1,3}^2-\frac{5}{16}g_{1,3}^4 -\frac{9}{8}g_{2,3}^2g_{1,3}^2-2h_3'^2-3h_3''^2 \nn \\
	&& +3g_{1,3}^2\lambda_{1,3}\Big) -\frac{1}{96}\Big(9L_b-3L_f-2\Big)\Big((N_c+1)g_2^4 + \frac{1}{6}Y_{2f}g_1^4\Big)n_f + \frac{N_c}{32}\Big(7L_b-L_f-2\Big)g_2^2y_t^2 \nn \\
	&& -\frac{N_c}{4}(3L_b+L_f)\lambda_1y_t^2 + \frac{N_c}{96}\Big(\Big(9(L_b-L_f)+4\Big)Y_{\phi}^2-2\Big(L_b-4L_f+3\Big)(Y_q^2+Y_u^2)\Big)g_1^2 y_t^2 \nn \\
	&& -\frac{N_cC_F}{6}\Big(L_b-4L_f+3\Big)g_s^2y_t^2+\frac{N_c}{24}\Big(3L_b-2(N_c-3)L_f\Big)y_t^4\Big],
	\eea
	with $C_F=\frac{N_c^2-1}{2N_c}=\frac{4}{3}$ and $c \sim -0.348723.$ is the fundamental quadratic Casimir of $SU(3)$ and $Y_{2f}=\frac{40}{3}, Y_u=\frac{4}{3}, Y_{\phi}=1, Y_l=-1, Y_e=-2,Y_q=\frac{1}{3}, N_c=3$. \\
	And the two-loop mass expressions for the mass paramter of the singlet and the triplet are given as:
	\bea
	m_{S,3}^2 & = & \frac{1}{2}\Big[2m_S^2(\Lambda)+ T^2 \Big(\frac{1}{3}\lambda_{hs}(\Lambda)+ \lambda_s(\Lambda)\Big) -\frac{L_b}{(4\pi)^2}\Big(18\kappa^2(\Lambda) + 4\lambda_{hs}m_h^2(\Lambda)+12 \lambda_s m_S^2(\Lambda)\Big)+ \frac{1}{(4\pi)^4}\Big[\frac{9(3+2L_b)}{2}(120\lambda_s \kappa^2) \nn \\
	&& + 756 L_b^2 \lambda_s \kappa^2 \Big] + \frac{1}{(4\pi)^2}\Big(2(3g_{2,3}^2+g_{1,3}^2)\lambda_{hs,3}-8\lambda_{hs,3}^2 
	-96\lambda_{s,3}^2\Big)\Big(c+\ln\Big(\frac{3T}{\Lambda_{3d}}\Big)\Big)+\frac{T^2}{(4\pi)^2}\Big[\frac{(2+3L_b)}{12}(3g_2^2+g_1^2)\lambda_{hs} \nn \\
	&&-L_b\Big(\Big(\lambda_h + \frac{7}{3}\lambda_{hs}+4\lambda_s\Big)\lambda_{hs}+36\lambda_s^2\Big)-\frac{N_c}{6}(3L_b-L_f)y_t^2\lambda_{hs}\Big]\Big] \nn, \\
	 m_{\Delta,3}^2 & = & \frac{1}{2}\Big[2m_\Delta^2 + T^2\Big(\frac{1}{3}\lambda_{h\Delta}(\Lambda)+\frac{5}{3}\lambda_\Delta(\Lambda)+\frac{1}{2}g_2^2(\Lambda)\Big)-\frac{1}{16\pi^2}\Big[-2(6g_2^2-20\lambda_\Delta)m_\Delta^2L_b + 4 m_h^2\lambda_{h\Delta}L_b + T^2 \Big(\big(\frac{71}{18}+\frac{2}{9}N_f\big)g_2^4 \nn \\
	 && +\frac{20}{3}\lambda_\Delta g_2^2 + \frac{1}{2}\lambda_{h\Delta}g_2^2  + \frac{1}{6}\lambda_{h\Delta}g_1^2 +L_b\Big(\frac{5}{12}g_2^4-3\lambda_{h\Delta}^2 -\frac{880}{12}\lambda_\Delta^2 + \frac{11}{4}\lambda_{h\Delta}g_2^2+\frac{1}{4}\lambda_{h\Delta}g_1^2+20 \lambda_\Delta g_2^2 -\frac{20}{3}\lambda_{h\Delta}\lambda_\Delta-2\lambda_{h\Delta}\lambda_h\Big)\nn \\
	 && +\Big(c+ \ln(\frac{3T}{\Lambda_{3d}})\Big)\Big(-8 \lambda_{h\Delta,3}^2 -160 \lambda_{\Delta,3}^2 + 2 \lambda_{h\Delta,3}(3g_{2,3}^2+ g_{1,3}^2)+80\lambda_{\Delta,3}g_{2,3}^2-3g_{2,3}^4+24 g_{2,3}^2\delta_3 -24 \delta_3^2 + 8 g_{2,3}^2\delta_3'-16 \delta_3 \delta_3'\nn\\
	 &&-16\delta_3'^2\Big)-L_f\Big(\lambda_{h\Delta}y_t^2 + \frac{2}{3}g_2^4 N_f\Big)+\ln(2)\Big(6\lambda_{h\Delta}y_t^2 + 4 g_2^4 N_f\Big)\Big)\Big]\Big].
	\eea
	The other parameters which are used in the above expressions are computed as follows:
	\bea
		g_{2,3}^2 & = & g_2^2(\Lambda)T\Big[1+\frac{g_2^2}{(4\pi)^2}\Big(\frac{44-N_d-2N_t}{6}L_b +\frac{2}{3}-\frac{4N_f}{3}L_f\Big)\Big],\\
	g_{1,3}^2 & = & g_1^2(\Lambda)T\Big[1+\frac{g_1^2}{(4\pi^2)}\Big(-\frac{N_d}{6}L_b-\frac{20 N_f}{9}L_f\Big)\Big], \\
	h_3 & = & \frac{g_2^2(\Lambda)T}{3}\Big(1+\frac{1}{(4\pi)^2}\Big[\Big(\frac{44-N_d-2N_t}{6}L_b+\frac{53}{6}-\frac{N_d}{3}-\frac{2N_t}{3}-\frac{4N_f}{3}(L_f-1)\Big)g_2^2+\frac{g_1^2}{2}-6y_t^2+12\lambda_1+8\lambda_{ht}\Big]\Big), \nn \\
	h_3' & = & \frac{g_1^2(\Lambda)T}{4}\Big(1+ \frac{1}{(4\pi)^2}\Big[\frac{3g_2^2}{2}+\Big(\frac{1}{2}-\frac{N_d}{6}(2+L_b)-\frac{20 N_f}{9}(L_f-1)\Big)g_1^2-\frac{34}{3}y_t^2+12\lambda_1\Big]\Big), \\
	h_3'' & = & \frac{g_2(\Lambda)g_1(\Lambda)T}{2}\Big(1+\frac{1}{(4\pi)^2}\Big[-\frac{5+N_d}{6}g_2^2 + \frac{3-N_d}{6}g_1^2 + L_b (\frac{44-N_d}{12}g_2^2-\frac{N_d}{12}g_1^2) \nn\\
	&& -N_f(L_f-1)\Big(\frac{2}{3}g_2^2 + \frac{10}{9}g_1^2\Big)+2y_t^2 + 4\lambda_1\Big]\Big),\\
	\delta_3 & = & \frac{1}{2}g_2^2(\Lambda)T\Big(1+\frac{1}{(4\pi)^2}\Big[\lambda_{ht}+8\lambda_t+g_2^2\Big(\frac{16-N_d-2N_t}{3}-\frac{4}{3}N_f(L_f-1)+L_b\frac{44-N_d-2N_t}{6}\Big)\Big]\Big),\\
	\delta_3' & = & -\frac{1}{2}g_2^2(\Lambda)T\Big(1+\frac{1}{(4\pi)^2}\Big[4\lambda_t + g_2^2\Big(-\frac{20+N_d+2N_t}{3}-\frac{4}{3}N_f(L_f-1)+L_b\frac{44-N_d-2N_t}{6}\Big)\Big]\Big).\nn \\
	\eea
	where, $N_d=1$, $N_t= 1$ and $N_f=3$ to identify the contributions from the SM Higgs doublet, the real triplet and the fermions, respectively.

\section{Transport equations}\label{trans}
The diffusion equations for the left-handed quarks are as follows;
 \bea
 (3\kappa_t + 3 \kappa_b)v_w \mu'_{q3}-(3K_{1,t}+3 K_{1,b})v'_{q3}-6\Gamma_y(\mu_{q3}+\mu_t)-6\Gamma_m(\mu_{q3}+\mu_t)-6\Gamma_{ss}[(2+ 9\kappa_t+\kappa_b)\mu_{q3}+(1-9\kappa_t)\mu_t]=0, \nn \\ 
  \eea
 where, the effect from the Higgs has been neglected, since it has relatively smaller effect on the final baryon asymmetry \cite{Fromme:2006wx} and 
  \bea
  -(K_{1,t}+K_{1,b})\mu'_{q3}+(K_{2,t}+K_{2,b})v_w v'_{q3}-\Big(\frac{K_{1,t}^2}{\kappa_t D_Q} + \frac{K_{1,b}^2}{\kappa_b D_Q}\Big)v_{q3}=K_{4,t}v_wm_t^2 \theta''_t+K_{5,t}v_w(m_t^2)'\theta'_t.
  \eea
Similarly, the transport equations for the right-handed top quark is;
\bea
3\kappa_t v_w \mu'_t -3K_{1,t}v'_t-6\Gamma_y(\mu_{q3}+\mu_t)-6\Gamma_m(\mu_{q3}+\mu_t)-3\Gamma_{ss}[(2+9\kappa_t+9\kappa_b)\mu_{q3}+(1-9\kappa_t)\mu_t]=0,
\eea
and
\bea
-K_{1,t}\mu'_t +K_{2,t}v_wv'_t-\frac{K_{1,t}^2}{\kappa_t D_Q}v_t =K_{4,t}v_w m_t^2 \theta'' +K_{5,t}v_w(m_t^2)'\theta_t',
\eea
where, primes denotes the derivative w.r.t. the $z$ coordinate perpendicular to the wall. The redefinition of chemical potential which is used in the above transport equations is $\mu_{q3}=\frac{\mu_{t,2}+\mu_{b,2}}{2}$, $\mu_{t^{c},2}=\mu_t$, now becomes $\mu_{L} = (1+2\kappa_t + 2 \kappa_b)\mu_{q3}-2 \kappa_t \mu_{t}$, where, $K_{m,j}$ and $\kappa_i$ are certain moments in momentum space as given below;
\bea
\expval{X}=\frac{\int_{}^{}d^3p X}{\int d^3p f'_{+}(m=0)},\\
f_{\pm}(m_i)=\frac{1}{e^{\beta}\sqrt{p^2+m_i^2}\pm 1},
\eea
where,
\bea
\kappa_i & = & \expval{f'_{\pm}(m_i)}, \\
K_{1,i} & = & \expval{\frac{p_z^2}{\sqrt{p^2+m_i^2}}f'_{\pm}(m_i)},\\
K_{2,i} & = & \expval{p_z^2 f'_{\pm}(m_i)}, \\
K_{3,i} & = & \expval{\frac{1}{2\sqrt{p^2+m_i^2}}f'_{\pm}(m_i)},\\
K_{4,i} & = & \expval{\frac{|p_z|}{2(p^2+m_i^2)}f'_{\pm}(m_i)},\\
K_{5,i} & = & \expval{\frac{|p_z| p^2}{2(p^2+m_i^2)}f'_{\pm}(m_i)}.
\eea
The momenta functions are normalized in such a way that $\kappa_i=2$ and 1 for massless bosons and fermions, respectively. The values used for the interaction rates and the quark diffusion constant are $\Gamma_{ws} = 1.0 \times 10^{-6} T^4$, $\Gamma_{ss}=4.9 \times 10^{-4}T^4$, $\Gamma_y=4.2 \times 10^{-3}T$, $\Gamma_{m}=\frac{m_t^2}{63 T}$, and $D_Q=\frac{6}{T}$ \cite{Huet:1995sh,Arnold:1999uy,Bodeker:1998hm,Moore:1997im}.  
\bibliography{References}
\bibliographystyle{Ref}

\end{document}